\documentclass[acmsmall]{acmart}

\AtBeginDocument{%
	}

\usepackage{xcolor}

\usepackage{listings}
\usepackage[utf8]{inputenc} 
\usepackage[T1]{fontenc}    
\usepackage{hyperref}       
\usepackage{url}            
\usepackage{amsfonts}       
\usepackage{microtype}      

\usepackage{algorithm}
\usepackage[noend]{algpseudocode}
\usepackage{amsmath}
\usepackage{multirow}
\usepackage{textcomp}
\usepackage{stfloats}
\usepackage{verbatim}
\usepackage{graphicx}

\usepackage{amssymb}
\usepackage{enumerate}
\usepackage{color, soul}
\usepackage{array}
\usepackage{enumitem}

\usepackage{booktabs}

\usepackage{pifont}
\usepackage{bbm}

\usepackage[final]{changes} %
\definechangesauthor[name={Reviewer 1}, color=magenta]{R1}
\definechangesauthor[name={Reviewer 2}, color=orange]{R2}
\definechangesauthor[name={Reviewer 3}, color=red]{R3}

\usepackage{diagbox}

\usepackage{tabularx,xspace,multicol,balance}

\newcommand{\eg}{\emph{e.g.,}\xspace}
\newcommand{\cf}{\emph{c.f.,}\xspace}

\newcommand{\first}{\textsf{(i)}\xspace}
\newcommand{\second}{\textsf{(ii)}\xspace}
\newcommand{\third}{\textsf{(iii)}\xspace}
\newcommand{\fourth}{\textsf{(iv)}\xspace}
\newcommand{\fifth}{\textsf{(v)}\xspace}
\newcommand{\fram}{\textrm{C.F.B}\xspace} 

\setcopyright{cc}
\setcctype{by}
\acmDOI{10.1145/3808184}
\acmYear{2026}
\acmJournal{PACMSE}
\acmVolume{3}
\acmNumber{FSE}
\acmArticle{FSE177}
\acmMonth{7}
\acmSubmissionID{fse26mainb-p1639-p}
\received{2025-09-12}
\received[accepted]{2026-03-24}

\usepackage[skip=2pt]{caption} 

\begin{document}
	
	\title{CrypFormBench: Benchmarking Formal Analysis Capability of Large Language Models for Cryptographic Schemes}
    \renewcommand{\shorttitle}{Benchmarking Formal Analysis Capability of Large Language Models for Cryptographic Schemes}
	

	\author{Zhaoxuan Li}
	\orcid{0000-0002-2195-0799}
	\affiliation{%
		\department{State Key Laboratory of Cyberspace Security Defense}
		\institution{Institute of Information Engineering, CAS}
		\city{Beijing}
		\country{China}
	}
	\email{lizhaoxuan@iie.ac.cn}
	
	\author{Qionglu Zhang}
	\orcid{0000-0002-0927-0678}
	\affiliation{%
		\department{State Key Laboratory of Cyberspace Security Defense}
		\institution{Institute of Information Engineering, CAS}
		\city{Beijing}
		\country{China}
	}
	\email{zhangqionglu@iie.ac.cn}
	
	\author{Hengyuan Liu}
	\orcid{0009-0004-9073-0601}
	\affiliation{%
		\department{State Key Laboratory of Cyberspace Security Defense}
		\institution{Institute of Information Engineering, CAS}
		\city{Beijing}
		\country{China}
	}
	\email{liuhengyuan@iie.ac.cn}
	
	\author{Xiaoyan Gu}
	\orcid{0000-0003-0673-0058}
	\affiliation{%
		\department{State Key Laboratory of Cyberspace Security Defense}
		\institution{Institute of Information Engineering, CAS}
		\city{Beijing}
		\country{China}
	}
	\email{guxiaoyan@iie.ac.cn}
	
	\author{Xianhui Lu}
	\orcid{0000-0001-7091-5810}
	\affiliation{%
		\department{State Key Laboratory of Cyberspace Security Defense}
		\institution{Institute of Information Engineering, CAS}
		\city{Beijing}
		\country{China}
	}
	\email{luxianhui@iie.ac.cn}
	
	\author{Hongbo Liu}
	\orcid{0009-0004-2444-2302}
	\affiliation{%
		\department{State Key Laboratory of Cyberspace Security Defense}
		\institution{Institute of Information Engineering, CAS}
		\city{Beijing}
		\country{China}
	}
	\email{liuhongbo@iie.ac.cn}
	
	\author{Bingzheng Wang}
	\orcid{0009-0005-9309-1478}
	\affiliation{%
		\department{State Key Laboratory of Cyberspace Security Defense}
		\institution{Institute of Information Engineering, CAS}
		\city{Beijing}
		\country{China}
	}
	\email{wangbingzheng@iie.ac.cn}
	
	\author{Haihui Fan}
	\orcid{0000-0002-4366-9121}
	\affiliation{%
		\department{State Key Laboratory of Cyberspace Security Defense}
		\institution{Institute of Information Engineering, CAS}
		\city{Beijing}
		\country{China}
	}
	\email{fanhaihui@iie.ac.cn}
	
	\author{Ziming Zhao}
	\orcid{0000-0003-1455-4330}
	\affiliation{%
		\institution{Zhejiang University}
		\city{Hangzhou}
		\country{China}
	}
	\email{zhaoziming@zju.edu.cn}
	
	\author{Rui Zhang}
	\orcid{0000-0003-0002-5593}
	\affiliation{%
		\department{State Key Laboratory of Cyberspace Security Defense}
		\institution{Institute of Information Engineering, CAS}
		\city{Beijing}
		\country{China}
	}
	\email{zhangrui@iie.ac.cn}
	
	\author{Li Zhou}
	\orcid{0000-0002-9868-8477}
	\affiliation{%
		\institution{Institute of Software, CAS}
		\city{Beijing}
		\country{China}
	}
	\email{zhouli@ios.ac.cn}

	\renewcommand{\shortauthors}{Z. Li, Q. Zhang, H. Liu, X. Gu, X. Lu, H. Liu, B. Wang, H. Fan, Z. Zhao, R. Zhang, and L. Zhou}
	
	
	\begin{abstract}
		Manual formal analysis of cryptographic schemes is labor-intensive and requires substantial expertise. While model-checking tools (\eg Scyther and Tamarin) and computational-security tools (\eg CryptoVerif and EasyCrypt) improve the automation of security proofs, they still rely on experts to abstract schemes and write tool-specific formal descriptions. Large language models (LLMs) are a promising alternative, but their effectiveness in this domain remains unexplored due to the absence of standardized evaluation methodologies. To fill this gap, we introduce CrypFormBench (\fram for short), a comprehensive benchmark jointly covering symbolic and computational security to evaluate five core LLM capabilities: interpretation, generation, completion, transformation, and correction. It comprises 700 instances spanning 677 schemes, 7 mainstream formal verifier languages, and 160 security properties. The evaluation of 9 state-of-the-art LLMs reveals that most of them perform well on interpretation and completion, given their code-awareness advantages, but struggle with generation, transformation, and correction. Overall, their performance remains limited, with Claude-3.5 achieving the highest score at 48.7 out of 100. We further provide practical guidance, \eg few-shot prompting, Pass@K sampling, and lightweight fine-tuning, to mitigate the executability bottleneck and improve tool-usable outputs. Taken together, our benchmark and analyses offer a grounded view of current progress and concrete directions toward reliable LLM-assisted formal cryptographic analysis.
	\end{abstract}


	
	\maketitle
	
	\section{Introduction}

Verifying the security of cryptographic schemes and primitives is a long-standing but challenging task in formal methods~\cite{dambrosi2024symbolic,TDSC_PSI,TCSS_PIWS}. Even widely deployed schemes such as TLS~\cite{dierks2008tls}, 5G-AKA~\cite{tsay20205g}, SSH~\cite{bellare1995ssh}, and OAuth 2.0~\cite{hardt2012oauth} have exhibited subtle flaws discovered years after deployment~\cite{needham1978using, lowe1995attack}. Automated formal methods provide a rigorous way to uncover such vulnerabilities before deployment. Existing approaches fall into two categories:  
\first model-checking tools (\eg Scyther~\cite{CAV_Scyther}, AVISPA~\cite{CAV_AVISPA}, ProVerif~\cite{blanchet2001proverif}, \added{and} Tamarin~\cite{CAV_TAMARIN}) for symbolic verification, and \second computational frameworks (\eg CryptoVerif~\cite{SP_CryptoVerif} \added{and} EasyCrypt~\cite{CRYPTO_EasyCrypt}) for cryptographic proof-level assurance.  

\begin{figure*}[t]
	\centering
	\includegraphics[width=0.87\textwidth]{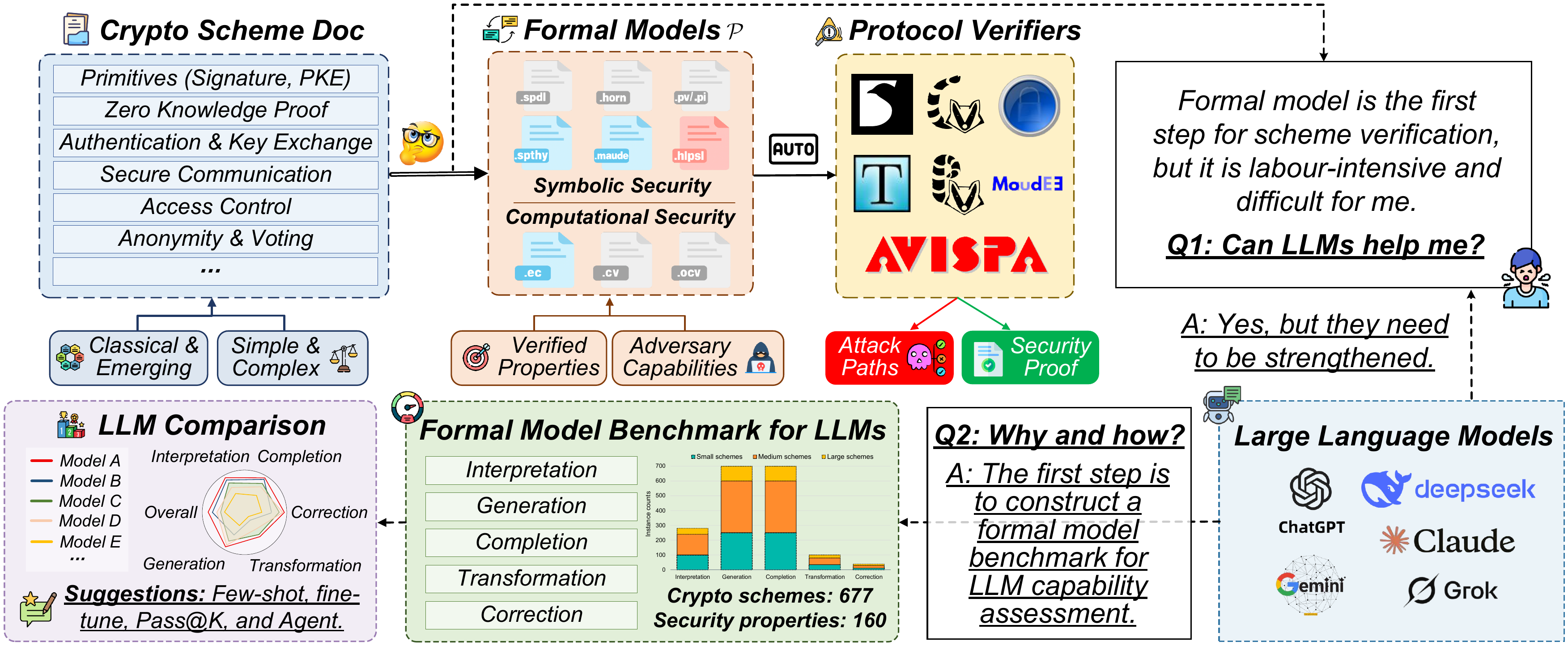}
	\caption{\added{Motivation for benchmarking LLM-assisted Cryptographic Scheme analysis.}}
	\label{fig:overview}
        \vspace{-0.3cm}
\end{figure*}

Nevertheless, these methods require substantial manual effort: abstracting scheme logic into specialized, error-prone, \added{and tool-compatible} languages (\eg HLPSL \added{and} applied $\pi$-calculus) demands expertise in both cryptography and formal specification. This bottleneck motivates exploring large language models (LLMs)~\cite{WWW_Carbon,WWW_HeteroSim} to assist in formal analysis by automatically deriving formal specifications from natural-language descriptions, as depicted in Fig.~\ref{fig:overview}. 
The current LLMs, particularly code-oriented variants, have shown promising capabilities in code generation~\cite{chen2021evaluating}, symbolic reasoning~\cite{bubeck2023sparks}, and even guiding proof assistants like Coq~\cite{Coq} and Lean~\cite{NeurIPS_LeanDojo, yang2023proofgpt, wu2024treeofthoughts}. Yet cryptographic scheme specifications remain a uniquely challenging domain due to their combination of strict syntax, security-critical semantics, and long contextual dependencies, where \replaced{small errors can break verification}{errors directly compromise verification}. Prior studies indicate that LLMs often struggle with such highly structured, domain-specific languages~\cite{dambrosi2024symbolic, gao2022pal}, and may \replaced{miss critical information}{fail to consistently attend to critical information} in long inputs~\cite{liu2023lost}. \replaced{Thus, their end-to-end formal capability for formal cryptographic analysis remains unclear.}{How well LLMs can handle end-to-end formal cryptographic analysis thus remains unclear.}

Recently, a preliminary benchmark~\cite{LLMAided} was proposed to evaluate automatic modeling tasks that support only symbolic security, with 18 real-world protocols and excluding computational security, advanced primitives, and diverse threat models. Unlike general-purpose code benchmarks~\cite{chen2021evaluating, hendrycks2021measuring}, constructing a benchmark for cryptographic scheme analysis poses unique challenges: complex semantics, multiple formal languages, and various adversarial assumptions. These gaps call for a more comprehensive and systematic evaluation framework.

To address these challenges, we design \textsc{CrypFormBench} (\fram for short), the first benchmark that jointly covers symbolic and computational security, providing a unified platform for evaluating LLMs in cryptographic scheme analysis. 
It spans five core tasks, namely \emph{interpretation}, \emph{generation}, \emph{completion}, \emph{transformation}, and \emph{correction}, reflecting the essential capabilities required for \emph{tool-usable} formal specification. 
It includes 700 curated instances (filtered from 4549 projects on GitHub or web sources) across seven verifiers' languages, covering 677 schemes and 160 security properties, from classical authentication scenarios to modern MPC, ZK, and post-quantum designs. 
Furthermore, we built an automated evaluation pipeline that standardizes model interactions, integrates cross-tool verification, and applies a multidimensional scoring framework that combines task-level metrics with overall rankings. This enables scalable and fine-grained assessment of LLM performance.  

In summary, we make the following main contributions. 
\begin{itemize}[leftmargin=2em, itemindent=0cm]
	\item We introduce \fram, a large-scale benchmark across seven verifier languages, enabling systematic evaluation of LLMs in cryptographic scheme analysis (\cf \S~\ref{sec:benchmark}).  
    \item We propose a unified evaluation methodology that integrates automated model interaction, multi-tool verification, and a weighted scoring framework (\cf \S~\ref{sec:testbed}).  
	\item We conduct the first comprehensive evaluation of nine state-of-the-art LLMs (\eg \textit{GPT-4o}, \textit{Claude-3.5}, and \textit{DeepSeek-Coder}), revealing strengths in interpretation/completion and weaknesses in generation/transformation/correction (\cf \S~\ref{sec:evaluation}), highlighting the challenges of domain-specific language generation. Also, we present robustness/optimization analyses and fine-grained property-level evaluation (\cf \S~\ref{sec:discussion}), demonstrating the stability of our findings. 
\end{itemize}

	\section{Preliminaries}

\subsection{\replaced[id=R2]{Formal Language of Cryptographic Scheme Analysis Tools}{Formal language of analysis tools}} \label{sec:tools}



Automatic cryptographic-scheme analysis is driven by the formal languages of verification tools, whose assumptions range from symbolic (Dolev-Yao) models to computational proofs under standard assumptions~\cite{SP_SoK}. Table~\ref{tab:tools} lists representative, widely adopted tools and languages spanning diverse techniques. Scyther~\cite{CAV_Scyther} and AVISPA~\cite{CAV_AVISPA} offer efficient and accessible analysis of classical authentication schemes. ProVerif~\cite{blanchet2001proverif} and Tamarin~\cite{CAV_TAMARIN} enable deep reasoning about equivalence and unbounded behaviors. Maude-NPA~\cite{FOSAD_Maude} supports algebraic reasoning. CryptoVerif~\cite{SP_CryptoVerif} and EasyCrypt~\cite{CRYPTO_EasyCrypt} provide computational guarantees aligned with security properties. As surveyed in~\cite{cortier2018survey}, these tools constitute the SOTAs for the formal analysis of cryptographic schemes. 

\begin{table*}[t]
\centering
\caption{Representative formal tools and their specification languages.}
\label{tab:tools}
\resizebox{0.85\linewidth}{!}{
\begin{tabular}{lll}
\toprule
\textbf{Tool} & \textbf{Language} & \textbf{Key Characteristics / Advantages} \\
\midrule
Scyther~\cite{CAV_Scyther} & SPDL & Lightweight and efficient symbolic model for secrecy/authentication claims. \\
Tamarin~\cite{CAV_TAMARIN} & SPTHY & Support stateful protocols and observational equivalence. \\
AVISPA~\cite{CAV_AVISPA} & HLPSL & Supports multiple parallelized back-end engines (OFMC, CL-AtSe, SATMC). \\
ProVerif~\cite{blanchet2001proverif} & PV/HORN/... & Handle unbounded sessions, and support for secrecy/equivalence properties. \\
Maude-NPA~\cite{FOSAD_Maude} & MAUDE & Analyze algebraic properties, and handle timeouts and internal errors. \\
CryptoVerif~\cite{SP_CryptoVerif} & CV/OCV & Computational soundness under both classical and random oracle models. \\
EasyCrypt~\cite{CRYPTO_EasyCrypt} & EC & Higher-order logic for computational proofs, and has machine-checked rigor. \\
\bottomrule
\end{tabular}}
\vspace{-0.3cm}
\end{table*}

\subsection{\replaced[id=R2]{A Generative Example for Automatic Formal Analysis}{A generative example for automatic analysis}}

Natural-language scheme descriptions can be mapped into formal languages required by analysis tools~\cite{CoRR_Lambda}. 
For example, ``the client sends a hash value of $n$ to the server'' can be expressed as:

SPDL (Scyther): send\_1(C, S, h(n));\quad \quad \quad
HLPSL (AVISPA): C $\rightarrow$ S : hash(n);

PV/HORN/... (ProVerif): out(c, hash(n));\quad 
SPTHY (Tamarin): [ Fr(n) ] --[ Out(hash(n)) ]-> []

Once generated, these specifications can be analyzed by their tools under the adversaries such as Dolev-Yao~\cite{cortier2018survey}. 
This enables a practical pipeline: LLMs translate natural-language descriptions into tool-compatible formal code, which is checked for properties such as secrecy/authentication~\cite{CAV_TAMARIN}. 

\textbf{Q1. How does formal code generation differ from traditional code generation?} 
Unlike traditional code generation, formal specifications require strict syntax and semantics~\cite{USENIX_SAPIC+}: small omissions (e.g., missing claims or wrong role bindings) can break compilation or yield misleading analysis outcomes~\cite{cortier2018survey}. 
Also, the same scheme often maps into multiple formal languages (\eg SPDL, HLPSL, PV, and SPTHY)~\cite{INDOCRYPT_Verifpal}, increasing manual effort and stressing cross-language generalization. 

\textbf{Q2. Can LLMs directly discover scheme flaws?}
Despite producing formal-looking specifications, LLMs cannot effectively verify schemes on their own, as this typically requires symbolic or computational security reasoning~\cite{LLMAided}. They often generate look-valid but incorrect results, and fail to capture freshness conditions, role bindings, or subtle attacks such as replay and man-in-the-middle~\cite{lowe1995attack}. 
Thus, LLMs are better for bootstrapping models, while verifiers must validate and refine the results, motivating systematic benchmarking to measure gaps and guide improvements. 

\textbf{\replaced{Q3. What makes benchmarking LLMs hard in formal analysis?}{Q3. What are the challenges in benchmarking formal analysis with LLMs?}} 
Nevertheless, building such a benchmark remains non-trivial. 
\first \textit{Dataset scarcity:} despite the abundance of cryptographic schemes, formal specifications are rarely available~\cite{dambrosi2024symbolic}, and \textit{scheme diversity} complicates unification. 
\second \textit{Cross-tool heterogeneity:} tools span different languages/paradigms, challenging integration~\cite{dierks2008tls}. 
\third \textit{Outcome variability:} tool outputs range from safe/unsafe to attack traces~\cite{CAV_TAMARIN,CAV_Scyther}, complicating automatic scoring. 
\fourth \textit{Evaluation at scale:} metrics must capture semantic soundness (beyond analyzability), while timeouts/explosions hinder scalability~\cite{LLMAided}. 
These challenges motivate a replicable benchmark for fair comparison and practical automation of formal cryptographic analysis.

\subsection{\added[id=R2]{Adversarial Assumptions}}
Security analysis critically depends on the underlying adversary model~\cite{cortier2018survey,lowe1995attack}. 
\first \textit{Passive DY} only observes traffic (no interception or injection), mainly used for secrecy analysis~\cite{blanchet2001proverif}.  
\second \textit{Active DY} controls the network (intercept/replay/forge/block), enabling checks of authentication and replay-style attacks~\cite{CAV_TAMARIN,CAV_Scyther}. 
\third \textit{Extended adversaries} capture domain-specific powers, \eg password-guessing in Internet-of-Things (IoT) or device-capture scenarios with key exposure~\cite{USENIX_SAPIC+,INDOCRYPT_Verifpal}. 

	\section{\fram Benchmark} \label{sec:benchmark}

\subsection{Corpus Overview}

\begin{figure*}[t]
	\centering
	\includegraphics[width=0.995\textwidth]{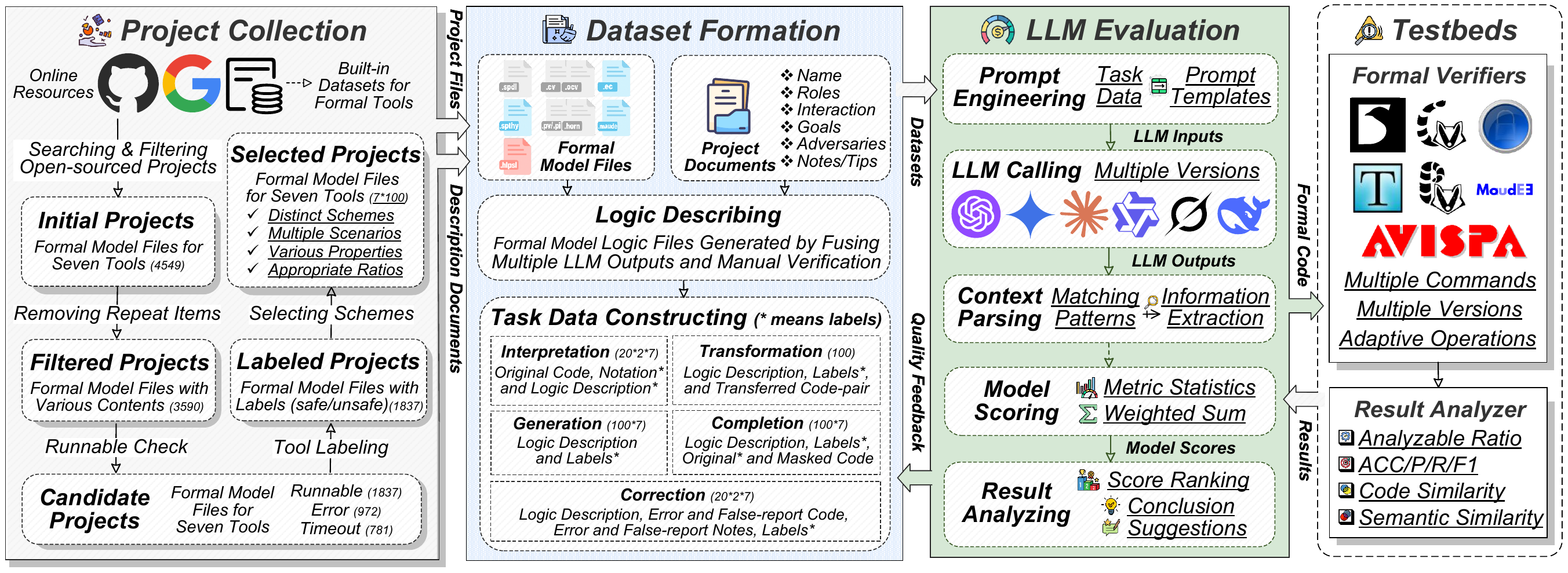}
	\caption{\added{The workflow of} \fram (\underline{C}rypto \underline{F}ormal \underline{B}enchmark).}
	\label{fig:work_flow}
	\vspace{-0.3cm}
\end{figure*}

Figure~\ref{fig:work_flow} depicts the end-to-end workflow of \fram for automatic LLM evaluation on formal cryptographic scheme analysis. Given a model, \fram runs unified tasks and outputs capability scores and overall rankings as follows. 
\first \textbf{Project collection.} We gather formal model files from open-source projects and tool-built datasets. After deduplication, executability checks, and tool-based labeling, we obtain a set of items that involve diverse schemes, scenarios, and security properties across the formal languages of seven tools, as detailed in Table~\ref{tab:tools}. 
\second \textbf{Dataset formation.} \added[id=R2]{We unify and verify formal code and scheme logic descriptions (LLM-assisted with manual checks),} then build task datasets (interpretation, generation, completion, transformation, and correction) with paired code, labels, and annotations. 
\third \textbf{LLM evaluation.} We query multiple LLMs with prompts, normalize/parse outputs, and verify them via formal tools  (\eg Scyther, Tamarin, and EasyCrypt) on \textbf{testbeds} to further compute metrics such as analyzability, accuracy, F1 score, and similarity. Five capability scores are aggregated into overall results and rankings. 











\subsection{Project \replaced{Collection}{collection}} \label{sec:collection}
\subsubsection{Pipeline}
The first stage of \fram is to prepare formal model files as the benchmark foundation in three steps:
\first \textbf{project sourcing.} 
We collect formal specification projects from open repositories (\eg GitHub/Google) and tool-provided datasets. This yields 4,549 model files across seven tools. After deduplication and basic format checks, 3,590 unique projects remain. 
\second \textbf{Project Labeling.}
We execute each file with its corresponding verifier and standardize outcomes as: \textit{safe} (1,280), \textit{unsafe} (557), \textit{error} (972), and \textit{timeout} (781). 
We treat \textit{safe}/\textit{unsafe} as analyzable runs and use them for benchmark construction. 
\third \textbf{Project Selection.}
We further curate a balanced subset with diverse schemes, communication scenarios, and security properties (\eg secrecy and authentication), while controlling the \textit{safe}/\textit{unsafe} ratio. 
Meanwhile, we assign a canonical security label to each selected scheme from the relevant cryptographic literature/standards. The \textit{unsafe} instances have established attacks (often with traces), whereas \textit{safe} instances have no known attacks (sometimes with proof sketches) under the specified threat model. 
Then, we align tool outputs with these labels, treat disagreements as modeling bugs, and discard instances that cannot be fixed. 
Finally, we sample 100 representative analyzable files per tool, yielding 700 reference files in total (\textit{safe} 415/\textit{unsafe} 285). 





\begin{figure*}[t]
	\centering
	\includegraphics[width=1.0\textwidth]{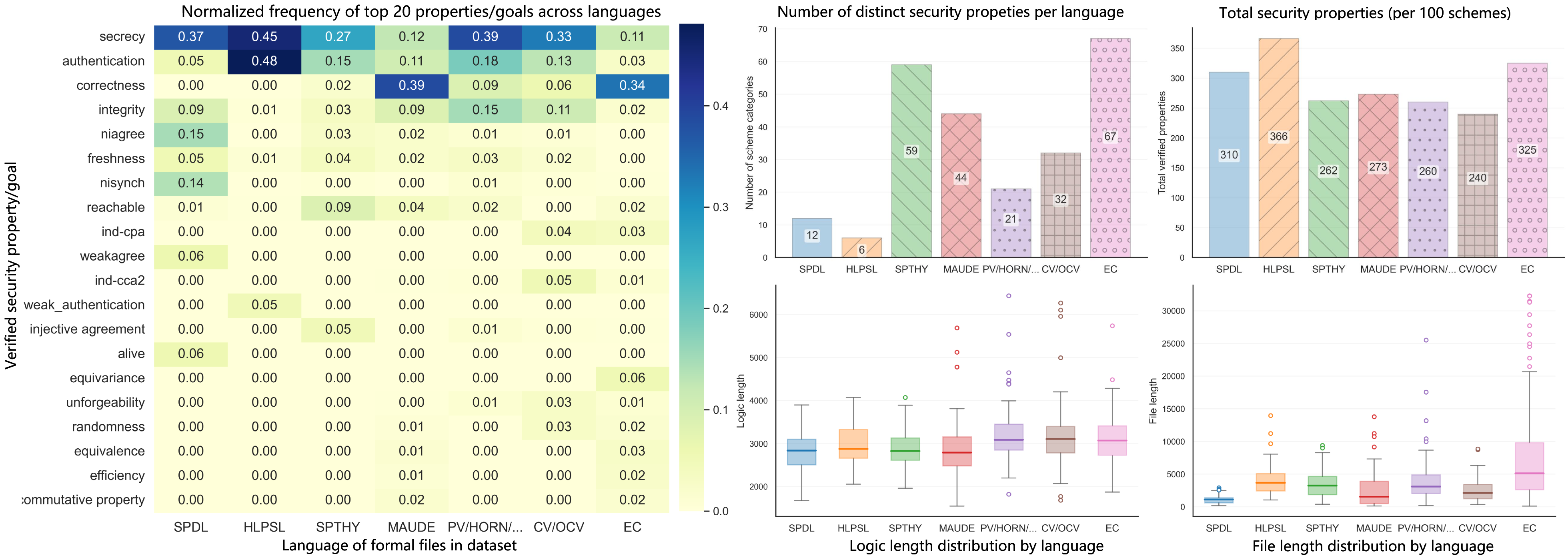}
	\caption{\added[id=R3]{The property and length distribution for various languages.}}
	\label{fig:property_distribution}
	\vspace{-0.3cm}
\end{figure*}

\subsubsection{Dataset Composition} \label{sec:dataset}
The 700 instances span seven formal languages and 677 distinct schemes across two-party/multiparty settings with diverse adversaries, forming a representative benchmark for both syntactic robustness and semantic correctness. 

\textit{Schemes.} 
The dataset includes classical and modern schemes for authentication, key exchange, zero-knowledge proofs, multiparty computation, and post-quantum designs, across domains such as Internet/IoT and blockchain. It also contains multi-language files of protocols such as \textit{Needham-Schroeder}~\cite{needham1978using}, \textit{Yahalom}~\cite{blanchet2001proverif}, and \textit{Diffie-Hellman}~\cite{lowe1995attack}. Their variants (\eg Paulson’s inductive and Lowe's corrected versions~\cite{lowe1995attack}) encode subtle differences that challenge fine-grained understanding. 

\textit{Security properties/goals.} 
We deduplicate 241 raw tool-specific goals into 160 distinct properties, covering secrecy, authentication, integrity, correctness, and equivalence.\footnote{\added[id=R3]{The full taxonomy of deduplicated properties refers to} \url{https://github.com/Secbrain/CrypFormBench/tree/main/datasets}.} 
As depicted in Fig.~\ref{fig:property_distribution}, although secrecy and authentication dominate, \fram still includes long-tail advanced goals (\eg forward secrecy, IND-CPA/CCA indistinguishability, and zero-knowledge). 
Each scheme has 2-3 verified properties on average, with complex cases (\eg multiparty key exchange) exceeding 10 properties.  
More importantly, different tools emphasize different properties. For instance, Scyther~\cite{CAV_Scyther} and ProVerif~\cite{blanchet2001proverif} focus on secrecy/authentication, while EasyCrypt~\cite{CRYPTO_EasyCrypt} highlights indistinguishability and reductionist proofs. 
This complementarity underscores the necessity of involving multiple tools in scheme analysis, motivates evaluating LLMs across diverse formal languages, and further calls for a unified score that enables fair cross-model comparison by summarizing end-to-end \emph{tool-usable} capability across heterogeneous verifiers/tasks under analyzability-correctness trade-offs. 

\textit{\added[id=R2]{Adversary models.}} 
The dataset \replaced{covers}{cover} Dolev-Yao, CK, eCK, and application-specific adversaries~\cite{cortier2018survey}, including replay, man-in-the-middle, KCI, and key-recovery attacks~\cite{lowe1995attack}. For example, \texttt{NAXOS\_eCK.spthy} models passive/active adversaries, while EasyCrypt~\cite{CRYPTO_EasyCrypt} and CryptoVerif~\cite{SP_CryptoVerif} encode computational assumptions such as CDH/DDH in the Random Oracle Model. This diversity ensures evaluation beyond syntax, capturing adversarial reasoning. 

\textit{Cross-language coverage.} 
A large portion of schemes in \fram are formalized in multiple verifier languages, involving the same tool family (\eg CV$\rightarrow$OCV) or across tools (\eg SPDL$\rightarrow$SPTHY), which enables controlled
cross-tool translation evaluation. 
For example, Needham-Schroeder \replaced{appears}{exists} in HORN, HORNTYPE, PI, and PV, while Diffie-Hellman is available in both PV and CV. 

\subsection{Dataset Formation}
\subsubsection{Logic Description}
\added[id=R2]{To enable systematic evaluation, each instance is paired with a unified logic description distilled from project documents, normalized across tools, and manually reviewed.} It consistently captures scheme roles, message flows, and security goals, bridging natural-language description narratives and tool-specific formal code. 

\begin{figure*}[t]
	\centering
	\includegraphics[width=0.90\textwidth]{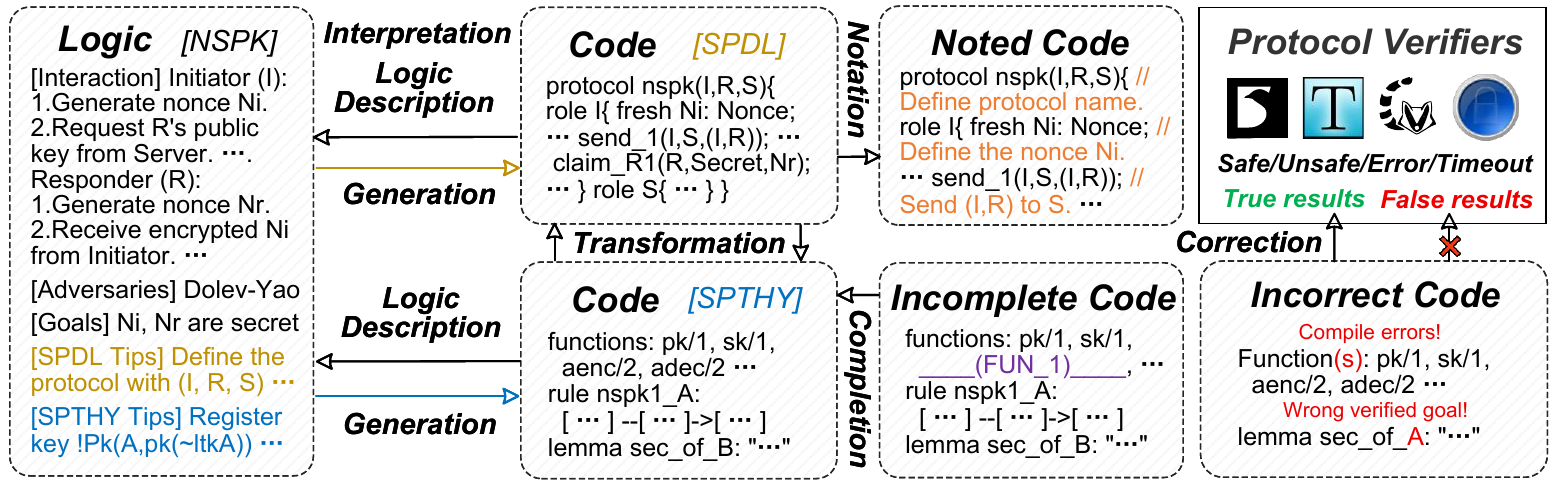}
	\caption{\added[id=R3]{Concrete examples for five benchmark tasks in} \fram.\protect\footnotemark \added[id=R1]{The tips only exist for the relevant language.}}
	\label{fig:examples}
	\vspace{-0.3cm}
\end{figure*}
\footnotetext{\added[id=R1]{The interactive and step-by-step examples are available on} \url{https://Secbrain.github.io/CrypFormBench}.}

\subsubsection{Task Data Construction}
Building on the logic descriptions, we design five tasks that probe complementary capabilities of LLMs in formal analysis, as an example shown in Fig.~\ref{fig:examples}. 
\first \textbf{Formal Code Interpretation (20*2*7 instances).}  
Given the original code, LLMs are asked to produce notation-level comments and global scheme summaries. Ground truth includes annotated notations and logic descriptions.  
\second \textbf{Formal Code Generation (100*7 instances).}  
From a natural-language logic description, LLMs generate complete formal code in the target language. 
Correctness is assessed by compiling the code and verifying whether the intended security goals are met in our testbed. 
\third \textbf{Formal Code Completion (100*7 instances).}  
Given partially masked specifications, LLMs fill in missing fragments. 
Masks are applied at different levels (\eg lemmas, type variables, statements) using grammar analyzers aligned with each tool’s syntax. 
Inputs include the masked code and logic description, while outputs are compared against the original files.
\fourth \textbf{Formal Code Transformation (100 instances).} 
LLMs are tasked with translating a scheme specification from one formal language to another for the same scheme and security properties under matched adversary assumptions, based on the source language and its logic description.\footnote{The transformation cases refer to \url{https://github.com/Secbrain/CrypFormBench/tree/main/datasets/transformation}.} A translation is considered correct if the verdict of transformed contents is consistent with that of the sources. 
\fifth \textbf{Formal Code Correction (20*2*7 instances).}  
LLMs repair code that either fails to compile or produces incorrect verification results. For compile errors, the inputs include the code, error message, and logic description, yet tips on wrong results are provided for false reports. Outputs are the corrected, analyzable specifications. 
Note that the instance counts are computed as a product: the first factor specifies the sample size (100 or 20), the second denotes the number of sub-tasks (if applicable), and the third indicates the number of languages (7, if any). 








\subsubsection{\added[id=R2]{Dataset Comparison}}

To emphasize the novelty of our dataset, we compare \fram with existing formal datasets of cryptographic schemes and LLM evaluation benchmarks, as shown in Table~\ref{tab:dataset-comparison}. 
\first \textbf{\added[id=R2]{Multi-language coverage.}} Seven mainstream formal languages are included, supporting both language-specific evaluation and cross-language transformation. 
\second \textbf{\added[id=R2]{Scale and diversity.}} 677 non-duplicated protocols cover classical (\eg Needham-Schroeder, Diffie-Hellman) and modern schemes (\eg MPC, ZK, post-quantum). 
\third \textbf{\added[id=R2]{Properties and adversaries.}} 160 distinct security properties and multiple attacker models (DY, CK, eCK, algebraic) extend well beyond tool demos. 
\fourth \textbf{\added[id=R2]{LLM orientation.}} Explicit labeling supports five tasks, making it the first dataset designed for systematic LLM evaluation in formal protocol analysis. 

\begin{table*}[t]
\centering
\caption{\added[id=R2]{Comparison of our dataset with existing benchmarks.}}
\label{tab:dataset-comparison}
\resizebox{0.89\linewidth}{!}{
\begin{tabular}{lccc}
\toprule
\textbf{Benchmark} & \textbf{Languages} & \textbf{Scales} & \textbf{Objects} \\
\midrule
AVISPA suite~\cite{CAV_AVISPA} & HLPSL (1) & $\sim$100 (Authentication, secrecy) & Tool-specific verification \\
Tamarin case studies~\cite{CAV_TAMARIN} & SPTHY (1) & $\sim$80 (Secrecy, agreement) & Tool-specific verification \\
ProVerif examples~\cite{blanchet2001proverif} & PV/... (1) & $\sim$60 (Secrecy, equivalence) & Tool-specific verification \\
EasyCrypt examples~\cite{CRYPTO_EasyCrypt} & EC (1) & $\sim$50 (Indistinguishability, reduction) & Computational proofs \\
LLM-aided~\cite{LLMAided} & SPTHY/PV (2) & 18 (Multi-party, secrecy, IoT) & Tool-specific verification  \\
\midrule
\textbf{\fram (ours)} & \textbf{SPDL/EC/...(7)} & \textbf{677} (Multi-party, 160 \replaced{properties}{propeties}, ZK, PQ, ...) & \textbf{Multi-task LLM evaluation} \\
\bottomrule
\end{tabular}}
\vspace{-0.3cm}
\end{table*}



%


	\section{End-to-End Evaluation Pipeline and Multi-Verifier Testbed} \label{sec:testbed}

\subsection{Input Preparation and LLM Interaction}
Each task is instantiated with a task-specific prompt constructed from our dataset. For example, in \emph{completion}, the prompt specifies the target language, provides an incomplete scheme file with placeholders (\eg \verb|```|\_\_\_\_()\_\_\_\_\verb|```|), and attaches its logic description. The model is explicitly instructed to fill only the missing parts, preserve all existing content, and ensure the output compiles in the designated tool (\eg Scyther). This template-based prompting enforces consistency across tasks/languages and matches realistic analyst workflows. 

\subsection{LLM Output Pre-processing}
Raw outputs of LLMs often contain formatting noise or inconsistent styles, so we apply a standardized pre-processing pipeline: 
\first \textbf{code extraction.} LLMs sometimes wrap formal code in various delimiters (\eg \verb|```|spdl\verb|```|, Markdown fences, or plain text). Only the relevant code segments are extracted, discarding commentary or formatting artifacts. 
\second \textbf{Description filtering.} For interpretation, keep only structured logic fields aligned with the reference template. 
\third \textbf{Normalization.} Canonicalize whitespace/indentation and variable aliases to avoid spurious mismatches. 
\fourth \textbf{Validity checks.} Filter empty or corrupted outputs before tool execution. 

\subsection{Testbed Construction and Verification}
To maximize coverage and reliability, the verification testbed integrates multiple tools, versions, and execution commands. 
\first \textbf{Multiple Tools.} As detailed in \S~\ref{sec:tools}, seven representative tools are assembled to cover a broad spectrum of security properties and adversarial models. 
\second \textbf{Multiple Versions.} For tools with unstable grammar across releases (\eg Maude-NPA v2.0/v3.0.1, CryptoVerif v1.25/v2.07, EasyCrypt 1.0/dev), we execute multiple versions from high to low. Stable tools such as Scyther, Tamarin, ProVerif, and AVISPA are run with a canonical version. 
\third \textbf{Multiple Commands.} To account for varying call behaviors, we employ alternative command sets. For instance, AVISPA invokes multiple engines (ofmc, cl-atse) sequentially, as a single engine may miss unsafe cases. Maude-NPA executes both ``red initials(0,12)'' and ``red initials(0,5)''. The former offers more precise results but risks timeout or internal errors, while the latter provides a fallback. Similarly, Tamarin enables observational equivalence via adaptive operation detection, and CryptoVerif automatically selects between classic and random-oracle models depending on file type (CV and OCV).  


\subsection{Evaluation Metrics} \label{sec:metrics}
Our framework evaluates model performance along five dimensions:
\first \textbf{Syntactic correctness.} Whether the generated code can be parsed and compiled by the testbed. The analyzability $a \in [0,1]$ refers to the analyzed ratio of generated formal files. 
\second \textbf{Functional correctness.} 
Whether the verified security properties (\eg secrecy and authentication) hold as intended, which is computed by matching scheme-level verdicts with the ground truth (SAFE/UNSAFE)\footnote{The ground-truth is supported by established analyses and aligns with the results of reference models, as detailed in \S~\ref{sec:collection}.}, as tools such as AVISPA provide only binary results. We compute the accuracy $\text{ACC}$ over all files, count the $\mathrm{TP}$/$\mathrm{TN}$/$\mathrm{FP}$/$\mathrm{FN}$ on analyzable files, and further calculate $\text{ACC}_\text{A}, \text{F1}_\text{A} \in [0,1]$. 
Also, the fine-grained property evaluation is discussed in \S~\ref{sec:properties}. 
\third \textbf{Execution efficiency.} Time required for verification, reflecting practical usability.  
\fourth \textbf{Code similarity.} Overlap between completed code and ground truth, used mainly in completion tasks. 
\fifth~\textbf{Semantic similarity.} Following prior work \cite{EMNLP_Sentence-BERT, ACL_Sentence, ICLR_code2seq, EMNLP_CodeBERT} on sentence-semantic similarity using vector representations, we quantify the quality of logic interpretation and code annotation by computing the cosine similarity $s_{\text{logic}}, s_{\text{anno}} \in [0,1]$ between the \replaced{embedding}{feature} vectors of model output and human-verified reference extracted by the \textit{Qwen3-Embedding-8B} model~\cite{Qwen3-Embedding}.

\subsection{Overall Capability Scoring Method} \label{sec:overall}
To enable a fair and unified evaluation of LLMs across the five capabilities tested in \fram, we design a comprehensive scoring scheme that integrates performance across tasks, languages, and LLMs. It balances \emph{executability} ($a$) and \emph{correctness} ($\text{ACC}_\text{A}$/$\text{F1}_\text{A}$ on analyzable outputs), with task-specific auxiliary signals such as code similarity (completion) and semantic similarity (interpretation). 

\paragraph{Executable tasks.} 
For each task $t$, language $L$, and model $M$, we compute metrics defined in \S\ref{sec:metrics} and test effectiveness via a two-level harmonic task score that penalizes imbalanced performance:
\begin{equation}
S_{\text{t}} = \mathrm{HM}(a^\gamma,\ Q), \quad Q = \mathrm{HM}(\text{ACC}_\text{A},\ \text{F1}_\text{A}), \quad
\mathrm{HM}(x,y)=\frac{2xy}{x+y+\varepsilon}, \quad \varepsilon=10^{-6},
\end{equation}
\added{where $\gamma$ is a tunable parameter (default $\gamma=1$) to penalize low analyzability.} 
On this basis, correction task consists of two sub-tasks: \textit{error correction} (syntactic/compilation errors) and \textit{false correction} (semantic mis-specifications). Their scores are combined as:
\begin{equation}
S_{\text{corr}} = \lambda_{\text{err}}\, S_{\text{err}} + \lambda_{\text{false}}\, S_{\text{false}},
\end{equation}
\added{where $\lambda_{\text{err}}=0.4$ and $\lambda_{\text{false}}=0.6$ (default), reflecting the greater difficulty of semantic repair.} 
\added[id=R3]{In addition to executability-based correctness on annotated code, interpretation also evaluates the semantic fidelity of the logic description and code annotations via embedding-based similarity. We combine them with default weights $\alpha=\beta=0.3$ by considering the importance of code availability.} 
\begin{equation} \label{eq:interpretation}
S_{\text{interp}} = \alpha\, s_{\text{logic}} + \beta\, s_{\text{anno}} + (1-\alpha-\beta)\,S_{\text{anno}}.
\end{equation}

\paragraph{Language aggregation.} 
\added{To avoid dominance of high-frequency languages, we adopt the macro-averaging method.} Optionally, difficulty weights $w_L$ can be applied to focus on specific languages. 
\begin{equation}
S_{t}(M)=\frac{1}{|{\mathcal L}|}\sum_{L\in{\mathcal L}} S_{t}(M,L).
\end{equation}

\paragraph{Task aggregation.} 
We assign task-level weights to reflect importance and difficulty.
\begin{equation}
S_{\text{overall}}(M) = \sum_{t \in \{\text{interp}, \text{gen}, \text{comp}, \text{trans}, \text{corr}\}} w_t\, S_t(M),
\end{equation}
where the default setting is $w_{\text{interp}}=0.15$, $w_{\text{gen}}=0.25$, $w_{\text{comp}}=0.20$, $w_{\text{trans}}=0.25$, and $w_{\text{corr}}=0.15$, motivated by task difficulty. 
For easier tasks (\eg interpretation and error correction), executability has fewer limitations. In contrast, small analyzability drops strongly affect usefulness for generation/transformation, so we upweight them. These weights are adjustable to match the evaluation focus and typically maintain stable ranking conclusions, which is discussed in \S~\ref{sec:parameters}.

	\section{Benchmark Results and Analysis} \label{sec:evaluation}

\subsection{Baseline Evaluation}
We comprehensively assess \fram with nine representative LLM baselines spanning proprietary and open-source, general-purpose and code- or reasoning-oriented models. 

\begin{itemize}[leftmargin=2em, itemindent=0cm]
	\item \textbf{GPT-4o}~\cite{gpt-4o} and \textbf{GPT-4o-mini}\cite{gpt-4o} (OpenAI): widely used proprietary models with strong reasoning and code understanding capabilities. Also, the latter follows a lightweight design. 
	
	\item \textbf{LLaMA4-Instruct}~\cite{llama4} (Meta): a representative open-source model for reproducibility. 
	
	\item \textbf{GLM-4}\cite{glm4} (ZhipuAI): a large-scale bilingual LLM optimized for reasoning and domain tasks. 
	
	\item \textbf{Gemini-2.5-Pro}\cite{gemini} (Google DeepMind): the latest frontier proprietary model with multi-modal and cross-lingual reasoning abilities, included to benchmark formal-spec performance. 
	
	\item \textbf{DeepSeek-R1}~\cite{deepseek-r1} and \textbf{DeepSeek-Coder}\cite{deepseek-coder}: a family of open-source models specialized in reasoning (R1) and programming/code generation (Coder).
	
	\item \textbf{Claude-3.5-Sonnet-Coder}~\cite{claude35sonnet} (Anthropic): a SOTA commercial model with enhanced code reasoning and safety alignment, suitable for formal specification tasks.
	
	\item \textbf{Grok-3}~\cite{grok}: an emerging commercial LLM focusing on reasoning-intensive applications. 
\end{itemize}


\textit{Experimental Setup.} All experiments and our testbeds were conducted on Ubuntu~22.04 with an AMD EPYC 9554P (64 cores), 512GB RAM, and 4$\times$NVIDIA L40 GPUs (192GB total). The \textit{Qwen3-Embedding-8B} \replaced{embedding encoder}{syntactic vector extraction model} and \textit{Qwen2.5-Coder-3B} model are run with \textit{transformers} library.

\begin{table*}[t]
\centering
\setlength\tabcolsep{2.0 pt}
\caption{Capability of LLMs evaluated on \fram according to the calculation method detailed in \S~\ref{sec:overall}.}
\resizebox{\textwidth}{!}{
\begin{tabular}{lcccccccccccccccccccccc}
\toprule
\multirow{2}{*}{Model} &
\multicolumn{4}{c}{Generation} & 
\multicolumn{4}{c}{Completion} &
\multicolumn{4}{c}{Correction} &
\multicolumn{4}{c}{Transformation} &
\multicolumn{4}{c}{Interpretation} &
\multirow{2}{*}{\begin{tabular}[c]{@{}c@{}}\textbf{$S_{\text{overall}}$}\end{tabular}}\\
\cmidrule(lr){2-5}\cmidrule(lr){6-9}\cmidrule(lr){10-13}\cmidrule(lr){14-17}\cmidrule(lr){18-21}
& $a$ & ACC$_\text{A}$ & F1$_\text{A}$ & \textbf{$S_{\text{gen}}$} 
& $a$ & ACC$_\text{A}$ & F1$_\text{A}$ & \textbf{$S_{\text{comp}}$}
& $a$ & ACC$_\text{A}$ & F1$_\text{A}$ & \textbf{$S_{\text{corr}}$}
& $a$ & ACC$_\text{A}$ & F1$_\text{A}$ & \textbf{$S_{\text{trans}}$}
& $a$ & ACC$_\text{A}$ & F1$_\text{A}$ & \textbf{$S_{\text{interp}}$} \\
\midrule
\multicolumn{22}{l}{\textit{Open-source Models}}\\
GLM-4
& 0.01 & 0.50 & 0.00 & \textbf{0.0}
& 0.13 & 0.94 & 0.92 & \textbf{22.8}
& 0.11 & 0.29 & 0.17 & \textbf{14.5}
& 0.05 & 0.00 & 0.00 & \textbf{0.0}
& 0.26 & 0.91 & 0.90 & \textbf{62.3}
& \textbf{16.1}
\\
LLaMA4-Instruct
& 0.03 & 0.79 & 0.77 & \textbf{5.8}
& 0.41 & 0.88 & 0.84 & \textbf{55.6}
& 0.56 & 0.59 & 0.58 & \textbf{57.2}
& 0.00 & 0.00 & 0.00 & \textbf{0.0}
& 0.15 & 0.95 & 0.94 & \textbf{44.4}
& \textbf{27.8}
\\
DeepSeek-Coder
& 0.03 & 0.85 & 0.80 & \textbf{5.8}
& 0.55 & 0.89 & 0.89 & \textbf{68.0}
& 0.61 & 0.67 & 0.67 & \textbf{63.9}
& 0.04 & 0.00 & 0.00 & \textbf{0.0}
& 0.86 & 1.00 & 1.00 & \textbf{93.2}
& \textbf{38.6}
\\
DeepSeek-R1
& 0.03 & 0.93 & 0.91 & \textbf{5.8}
& 0.50 & 0.94 & 0.93 & \textbf{65.2}
& 0.50 & 0.67 & 0.65 & \textbf{56.9}
& 0.08 & 0.00 & 0.00 & \textbf{0.0}
& 0.41 & 1.00 & 1.00 & \textbf{77.4}
& \textbf{34.6}
\\
\midrule
\multicolumn{22}{l}{\textit{Closed-source Models}}\\
GPT-4o-mini
& 0.06 & 0.00 & 0.00 & \textbf{0.0}
& 0.06 & 0.64 & 0.53 & \textbf{10.9}
& 0.41 & 0.51 & 0.48 & \textbf{44.8}
& 0.04 & 0.00 & 0.00 & \textbf{0.0}
& 0.79 & 1.00 & 1.00 & \textbf{93.8}
& \textbf{23.0}
\\
Gemini-2.5-Pro
& 0.04 & 1.00 & 1.00 & \textbf{7.7}
& 0.15 & 0.94 & 0.94 & \textbf{25.9}
& 0.22 & 0.93 & 0.93 & \textbf{35.6}
& 0.03 & 0.00 & 0.00 & \textbf{0.0}
& 0.34 & 0.89 & 0.90 & \textbf{69.1}
& \textbf{22.8}
\\
GPT-4o
& 0.05 & 1.00 & 1.00 & \textbf{9.5}
& 0.36 & 0.91 & 0.90 & \textbf{51.5}
& 0.48 & 0.65 & 0.62 & \textbf{54.7}
& 0.07 & 0.00 & 0.00 & \textbf{0.0}
& 0.81 & 1.00 & 1.00 & \textbf{94.1}
& \textbf{35.0}
\\
Claude-3.5-Sonnet-Coder
& 0.15 & 0.62 & 0.63 & \textbf{24.2}
& 0.62 & 0.93 & 0.93 & \textbf{74.4}
& 0.67 & 0.75 & 0.77 & \textbf{71.2}
& 0.14 & 0.07 & 0.62 & \textbf{13.3}
& 0.78 & 0.99 & 0.99 & \textbf{91.7}
& \textbf{48.7}
\\
Grok-3
& 0.07 & 0.74 & 0.71 & \textbf{12.8}
& 0.23 & 0.91 & 0.90 & \textbf{36.7}
& 0.12 & 0.50 & 0.00 & \textbf{0.0}
& 0.17 & 0.00 & 0.00 & \textbf{0.0}
& 0.01 & 0.00 & 0.00 & \textbf{49.6}
& \textbf{18.0}
\\
\bottomrule
\end{tabular}}
\label{tab_overall_results}
\vspace{-0.3cm}
\end{table*}

\subsection{\replaced[id=R3]{The Overall Capability}{The overall capability}}

In this section, we present an integrated discussion of the five key capabilities evaluated in \fram: generation, completion, correction, transformation, and interpretation. Table~\ref{tab_overall_results} \added[id=R3]{summarizes the performance of the evaluated models in terms of overall and task levels.} Several insights stand out: \first \textit{Claude-3.5-Sonnet-Coder} consistently ranks highest across all tasks, demonstrating strong robustness in both syntactic validity and semantic reasoning.
\second \textit{DeepSeek-Coder} excels in error correction and completion, and occasionally matches top models in interpretation and generation. 
\third \textit{GPT-4o} performs strongly in interpretation but struggles in generation and transformation, while its smaller variant \textit{GPT-4o-mini} is less stable yet competitive in some simpler languages.
\fourth General-purpose models such as \textit{Gemini-2.5-Pro} and \textit{GLM-4} generally underperform, highlighting the importance of code- and reasoning-oriented optimization.
\fifth Weaker baselines (\textit{LLaMA4-Instruct} and \textit{Grok-3}) consistently fail across most tasks, rarely producing analyzable outputs.
These complementary strengths suggest an \textit{agentic} workflow that routes tasks to the most suitable model. 

A cross-task comparison highlights clear patterns. Interpretation and completion are generally easier for most models, yielding a higher analyzed ratio and better accuracy. By contrast, transformation and generation remain the most difficult tasks: even the strongest models achieve only low analyzed ratio and limited correctness, underscoring the persistent challenges in these tasks. 

Overall, LLMs show tractable performance on interpretation and error correction, but code generation, especially cross-language transformation, remains a major bottleneck. Outcomes are shaped by model specialization and language complexity: domain-optimized models such as \textit{Claude-3.5-Sonnet-Coder} and \textit{DeepSeek-Coder} lead, yet reliable end-to-end automation remains unresolved. 


\vspace{2mm}
\noindent\fbox{%
	\parbox{0.97\columnwidth}{%
		\textbf{Can current LLMs provide end-to-end automation for formal scheme analysis?} LLMs can be regarded as strong assistants for interpretation and localized correction, but not yet reliable for full generation or cross-language transformation. 
		Future work should prioritize grammar-constrained decoding, domain-specific fine-tuning, agentic model orchestration, and tool-in-the-loop workflows to bridge the gap toward practical end-to-end automation. 
	}
}

\subsection{\replaced{The Generation Capability}{The generation capability}}


\begin{figure*}[t] 
	\centering
	\includegraphics[width=0.995\textwidth]{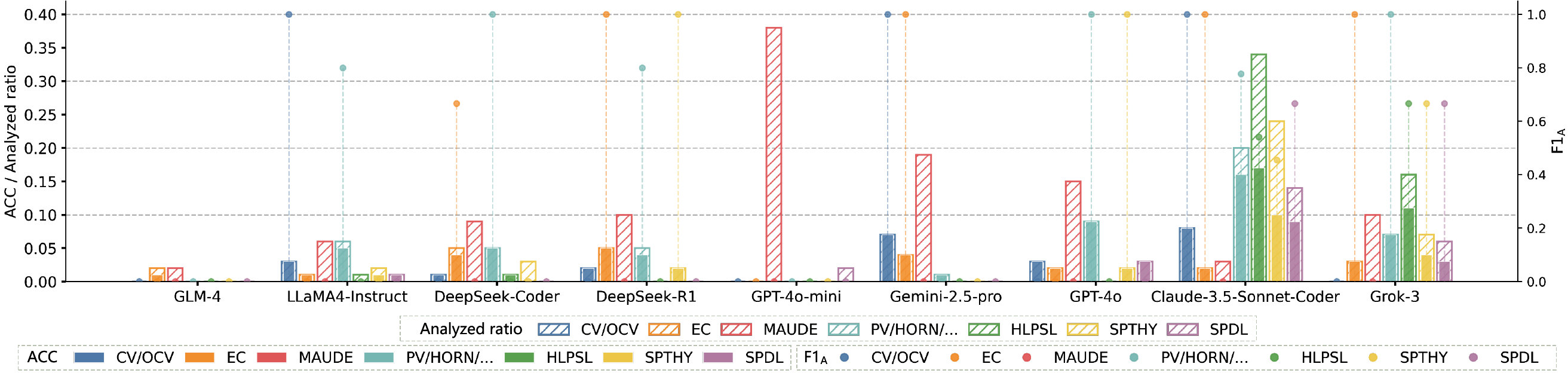}
	\caption{\added[id=R1]{Results of the generation experiment. Different colors represent different formal languages. Among them, light bars represent the analyzed ratio, dark colors represent ACC, and scattered dots represent $ \text{F1}_\text{A} $.}}
	\label{fig:generation}
	\vspace{-10pt}  
\end{figure*}

We compare LLM code generation across seven formal languages using three metrics: analyzed ratio, ACC, and $\text{F1}_\text{A}$, as shown in Fig.~\ref{fig:generation}. We observe four consistent patterns: \first closed-source models lead overall. \textit{Claude-3.5-Sonnet-Coder}, \textit{GPT-4o}, and \textit{Grok-3} follow prompts more reliably and achieve higher analyzability, verification success, and $\text{F1}_\text{A}$ than most open-source models, indicating stronger end-to-end usability. 
\second Strong intra-model variability across languages. Performance differs sharply by language within the same model: \textit{Gemini-2.5-Pro} is stronger on CV/EC (computational-security-oriented), \textit{GPT-4o} performs best on PV, while \textit{Claude-3.5-Sonnet-Coder} degrades on EC and Maude. This likely reflects uneven training coverage and language-specific grammatical/semantic complexity. 
\third Stable language-level trends across models. Except for \textit{GLM-4} and \textit{GPT-4o-mini}, most models obtain a higher analyzed ratio and ACC on PV than on SPTHY/SPDL/Maude, consistent with PV being closer to mainstream programming syntax (\eg Java/C). Maude is a broad failure mode: some outputs compile, but none satisfy target properties, suggesting sparse data and stronger contextual dependencies. 
\fourth Reasoning/coder models excel on computational proof languages (CV/EC). Reasoning/coder-oriented models (\textit{Gemini-2.5-Pro}, \textit{DeepSeek-R1}, \textit{Claude-3.5-Sonnet-Coder}, and \textit{DeepSeek-Coder}) consistently achieve a higher analyzed ratio, ACC, and $\text{F1}_\text{A}$ on CV/EC, showing advantages in structured generation and computational reasoning. 

In sum, performance varies across languages due to scarce/uneven data, language complexity, and model biases, making generation less reliable than for mainstream programming (\eg Python).

\vspace{2mm}
\noindent\fbox{%
	\parbox{0.97\columnwidth}{%
		\textbf{Question. Can LLMs directly generate analyzable and correct formal scheme models?} Current LLMs often fail to satisfy strict formal syntax/semantics, especially in complex languages. such as EC and PV/HORN. Practical improvement likely requires grammar-constrained decoding and tool-guided feedback/repair to reduce invalid outputs and improve usability.
	}
}

\subsection{\replaced{The Completion Capability}{The completion capability}}


\begin{figure*}[t] 
	\centering 
	\includegraphics[width=0.95\textwidth]{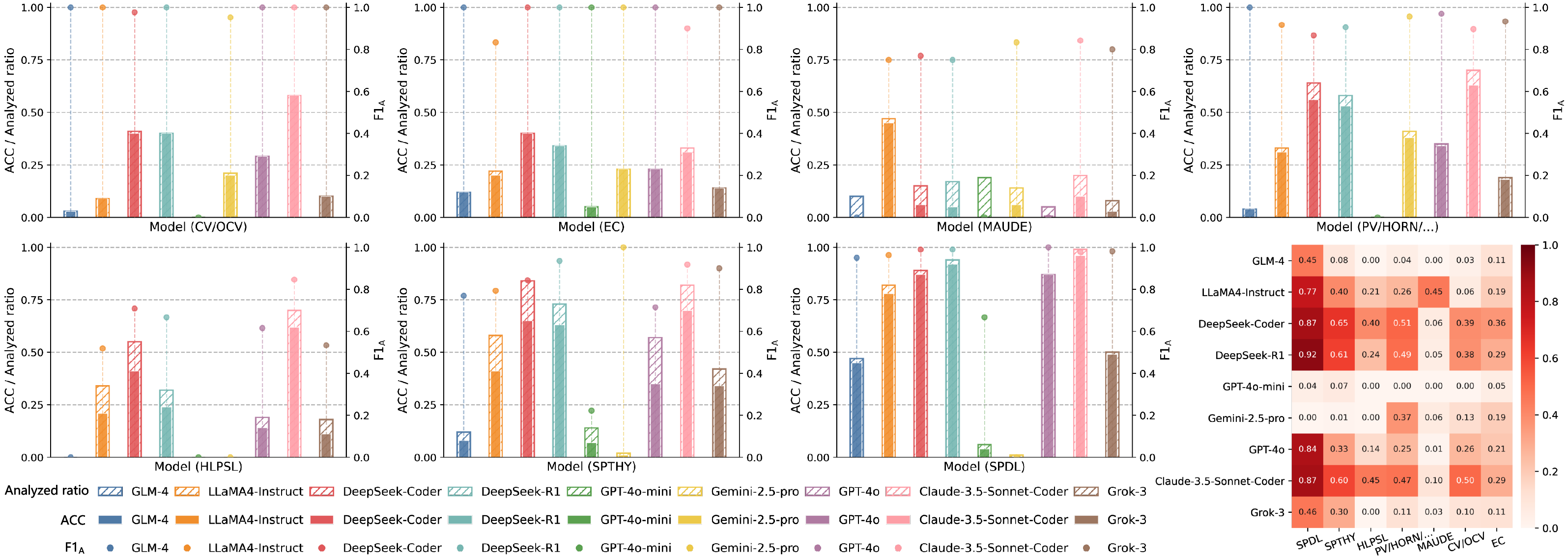}
	\caption{\replaced[id=R1]{Results of completion experiment. The last sub-figure shows ACC improvement of completion over generation across languages for each model. Other sub-figures report per-language performance: light bars indicate analyzed ratio, dark bars denote ACC, and scatter points show $\text{F1}_\text{A}$.}{Results of completion experiment. The last sub-figure demonstrates the improvement in ACC for the completion task compared to the generation task across different languages for each model. Others illustrate the performance of different models across various formal languages, where light shades represent the analyzed ratio, dark shades denote the ACC, and scatter points indicate the $\text{F1}_\text{A}$.}}
	\label{fig:completion}
	\vspace{-0.3cm}
\end{figure*}

Figure~\ref{fig:completion} reports the completion task, where models fill missing segments of formal code. Compared with full generation, completion provides a partially correct context, reducing the burden of producing valid syntax/semantics from scratch. We report the analyzed ratio, ACC, and $\text{F1}_\text{A}$.


\textit{Overall performance.}
Most models improve ACC across languages in the completion setting, confirming that partial context alleviates syntactic/structural burdens compared to generation. Notably, weaker models (\eg \textit{GPT-4o-mini}) break the near-zero barrier on EC/SPDL/SPTHY. For Maude, stronger models (\eg \textit{Claude-3.5-Sonnet-Coder}) can complete and verify a subset of cases. We further observe an all-or-nothing pattern: once outputs become analyzable, correctness is typically high (notably on EC/CV), making executability/analyzability the primary bottleneck. 



\textit{Language comparison.}
Most models (except \textit{GPT-4o-mini} and \textit{Gemini-2.5-Pro}) perform best on SPDL, likely due to its shorter instances and constrained, redundant syntax that provides stronger completion cues. Performance drops on complex languages: Maude and EC remain challenging, as in generation. \textit{LLaMA-Instruct} is an exception on Maude, plausibly due to greater pretraining exposure, while EC gains are smaller due to longer code and larger masked fragments.  



\textit{Model comparison.}
Code-specialized models consistently outperform general-purpose LLMs. \textit{Claude-3.5-Sonnet-Coder} achieves the strongest overall results with both high analyzed ratio and ACC. \textit{DeepSeek-R1}/\textit{DeepSeek-Coder} also improve substantially, reaching ACC above $30\%$ on most languages (while $<5\%$ in generation). In contrast, \textit{GPT-4o-mini} and \textit{GLM-4} pass only a limited number of cases, reflecting limited formal-code capability.

Overall, completion is easier than full generation, and context boosts analyzability, but producing syntactically valid code for complex formal languages still limits reliability.

\vspace{2mm}
\noindent\fbox{%
	\parbox{0.97\columnwidth}{%
		\textbf{Can LLMs reliably complete partial scheme specifications into analyzable formal models?} \replaced[id=R2]{Context}{Conetext} substantially \replaced{improves}{improve} analyzability, and once code is valid, correctness is typically high. 
		However, syntactic validity in complex languages (\eg Maude and EC) remains the main obstacle, motivating grammar-aware or tool-guided completion strategies. 
	}
}

\subsection{\replaced{The Correction Capability}{The correction capability}}



We evaluate correction capability under \first \textit{correction error}, where the specification fails to compile, and the model fixes it by handling compiler errors and the intended logic. \second \textit{correction false}, where the code compiles but violates expected properties, requiring semantic repair guided by the logic. 

\begin{figure*}[t]
	\centering
	\includegraphics[width=0.995\textwidth]{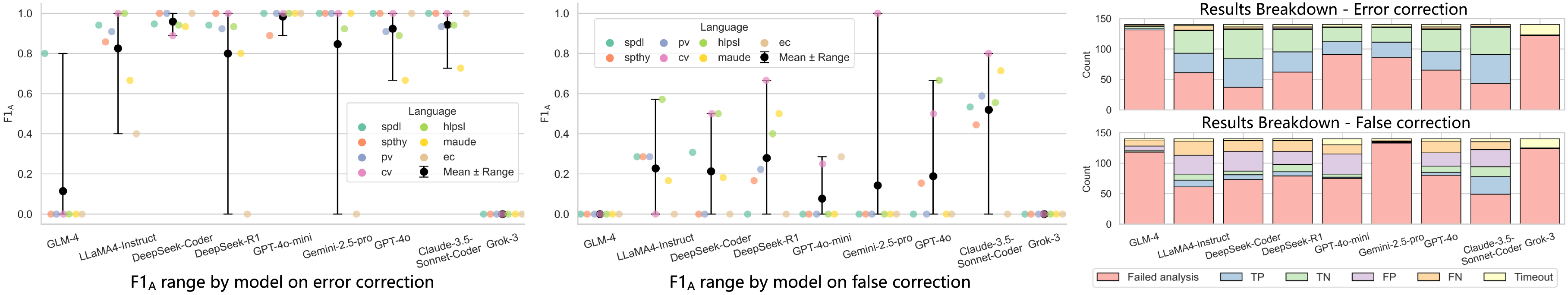}
	\caption{\added{Results of the correction experiment.} The left two show the $\text{F1}_\text{A}$ range per model on correction error and false, with colored dots denoting languages, and black markers indicating the mean $\pm$ range. The right stacked bars summarize categorical outcomes aggregated over all languages.} 
	\label{fig:correction}
	\vspace{-0.3cm}
\end{figure*}

\textit{Error correction.} 
Figure~\ref{fig:correction} shows a clear difficulty gap. In \textit{correction error}, models benefit from explicit diagnostics and typically localized syntactic issues. Code-specialized models (\eg \textit{Claude-3.5-Sonnet-Coder} and \textit{DeepSeek-Coder}) achieve an analyzed ratio around $0.6$ across languages, benefiting from the compiler feedback and formal-syntax adherence. Performance degrades on complex languages (\eg EC/Maude), where multiple interdependent fixes are required. Moreover, once outputs are analyzable, most models achieve high $\text{F1}_\text{A}$ (often $>0.9$), indicating that compiler-guided structural repair is largely within current LLM capabilities.


\textit{False correction.} 
This task is harder than error correction: although faulty code and logic are provided, models identify subtle semantic faults (\eg mis-specified properties or missing adversary actions) while preserving syntactic validity. 
Most models have low analyzed ratios and modest $\text{ACC}_\text{A}$. \textit{Claude-3.5-Sonnet-Coder} performs well, yet still trails error correction. 

Overall, LLMs can reliably fix localized, compiler-reported errors, but semantic-level debugging remains difficult, motivating tool-guided repair and human-in-the-loop validation.  


\vspace{2mm}
\noindent\fbox{%
	\parbox{0.97\columnwidth}{%
		\textbf{Question. Can LLMs reliably correct syntactic and semantic errors in formal models?} 
        Compiler-guided fixes are largely reliable and achieve high accuracy once analyzable, but semantic-level false corrections remain challenging, motivating tool feedback/human validation.
	}
}

\begin{figure*}[t]
	\centering
	\includegraphics[width=0.995\textwidth]{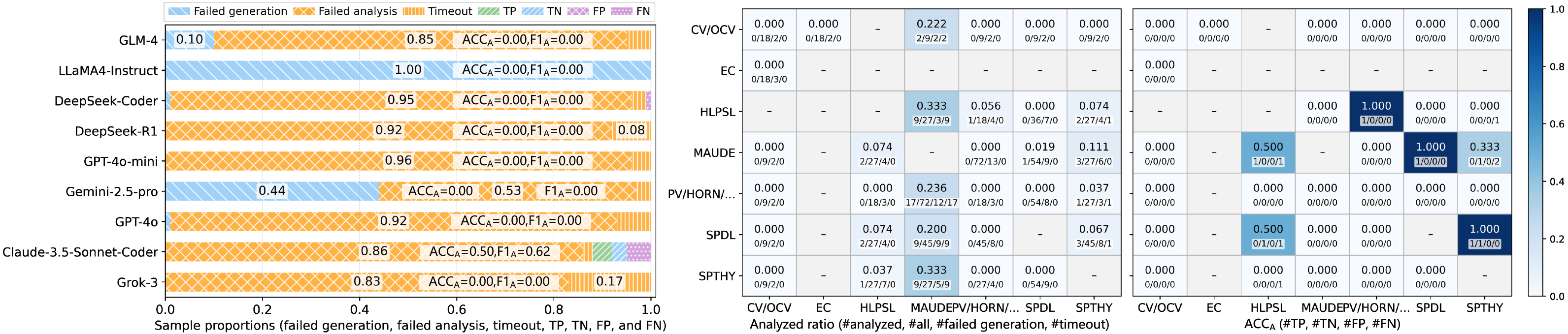}
	\caption{\added[id=R2]{Results of transformation experiment from model and language perspectives. The source$\rightarrow$target direction corresponds to y$\rightarrow$x in heatmaps, and the symbol ``-'' refers to no transformation instances.}}
	\label{tab_translation}
	\vspace{-0.35cm}
\end{figure*}

\subsection{\replaced{The Transformation Capability}{The transformation capability}}

Figure~\ref{tab_translation} summarizes the transformation results. At the model level (left), we break down outcomes into failed generation, failed analysis, timeout, and the TP/TN/FP/FN on analyzable cases. At the language level, we report the directed source$\rightarrow$target analyzable ratio and the corresponding $\mathrm{ACC}_\mathrm{A}$. 


\textit{\added[id=R2]{Cross-language interference and timeout bottleneck.}} 
A major failure mode is cross-language interference: given both source code and logic, models often copy source fragments into the target file, but equivalent operations use incompatible idioms across tools (\eg \texttt{Out(msg)} in SPTHY and \texttt{send\_1(A,B,msg)} in SPDL), causing early parse/type failures. Even among compilable translations, timeouts are frequent, indicating a semantic/structural mismatch. Manual inspection confirms that compiled outputs are often redundant or ill-structured: in rewriting-based tools (\eg Maude-NPA), unnecessary nesting (\eg \texttt{enc(enc(a,k1),k2)}) can trigger exponential rewriting/non-termination, while missing mandatory clauses (\eg \texttt{::nil::[+(null),nil]\&} in Maude) can also induce infinite loops. 
Additional recurring errors are summarized in \S~\ref{sec:cases}. The middle/right heatmaps further show that only a few source$\rightarrow$target directions achieve non-zero analyzability, and even fewer preserve correctness (non-trivial $\mathrm{ACC}_\mathrm{A}$), suggesting the core difficulty is robust cross-tool semantic mapping across heterogeneous formalisms, not just ``hard languages'' in isolation. 

\textit{Superiority of specialized models.} Among all LLMs, \textit{Claude-3.5-Sonnet-Coder} achieves the highest analyzed ratio (14\%) and meaningful correctness on analyzable cases (\eg $\text{F1}_\text{A}=61.54\%$), suggesting that coding-oriented models retain limited but promising cross-tool transformation ability.

In sum, translation is the most challenging setting because models must align source semantics with target syntax, causing only a small fraction of analyzable outputs (typically <10\%), even for powerful models (\eg \textit{GPT-4o} and \textit{DeepSeek-Coder}). We further discuss practical directions such as grammar-constrained decoding and tool-in-the-loop feedback in \S~\ref{sec:discussion}. 




\vspace{2mm}
\noindent\fbox{%
	\parbox{0.97\columnwidth}{%
		\textbf{Question. Can LLMs directly transform scheme specifications across different formal languages?} Analyzable outputs are scarce, with dominant failures from cross-language interference, syntactic incompatibility, and semantic/structural mismatch (timeouts/incorrect goals). Promising directions include grammar-constrained decoding to align semantics across tools.
	}
}

\subsection{\replaced{The Interpretation Capability}{The interpretation capability}}

\begin{figure*}[t]
	\centering
	\includegraphics[width=0.995\textwidth]{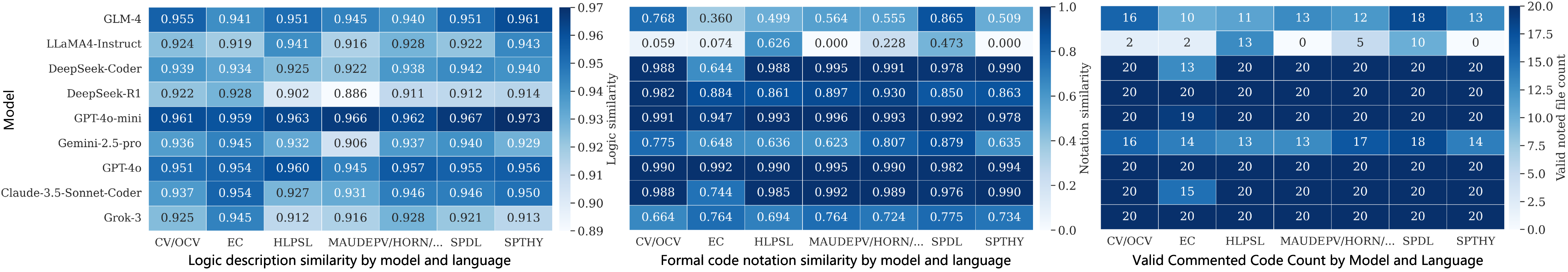}
	\caption{\added{Interpretation results across models and languages.} The left heatmap shows the similarities between the logic description and the intended logic. The middle reports annotation similarity with ground-truth explanations. \added[id=R3]{The right presents the count of valid commented code blocks.}}
	\label{fig:interpretation}
	\vspace{-0.35cm}
\end{figure*}




The evaluation of interpretation capability is summarized in Figure~\ref{fig:interpretation}. We assess two aspects: \first \textit{logic illustration}, measuring semantic similarity between generated logic description and ground-truth references, and \second \textit{code annotation}, evaluating both notation similarity and code executability.

\textit{Logic interpretation.}
Most models reconstruct interaction logic with high similarity. In the left heatmap, \textit{GPT-4o} and \textit{Claude-3.5-Sonnet-Coder} consistently exceed $0.95$ on average, while \textit{DeepSeek-R1} and \textit{DeepSeek-Coder} remain competitive. In contrast, general-purpose or lightweight models \textit{LLaMA4-Instruct} and \textit{Grok-3} struggle to capture subtle adversary behaviors and security goals. 



\textit{\added[id=R3]{Code annotation and executability.}} 
The middle/right heatmaps show that leading proprietary models (\textit{GPT-4o}, \textit{GPT-4o-mini}, and \textit{Claude-3.5-Sonnet-Coder}) produce highly similar comments (typically $>0.98$) while preserving tool-required syntax. In contrast, \textit{Gemini-2.5-Pro} and \textit{GLM-4} exhibit a gap: despite moderate comment similarity, their annotated files often fail to compile or execute, showing that fluent explanations do not guarantee tool compatibility. \textit{LLaMA4-Instruct} and \textit{Grok-3} underperform on both similarity and executability (near-zero executable outputs). 

Leading proprietary models produce highly similar comments (typically $>0.98$) while preserving tool-required syntax. \textit{Gemini-2.5-Pro} and \textit{GLM-4} show that fluent comments do not ensure compilability/executability, and \textit{LLaMA4-Instruct}/\textit{Grok-3} underperform on both.



Overall, executability is a necessary complement to similarity-based evaluation: annotations must be semantically faithful and satisfy strict formal-tool constraints. Thus, current LLMs are reliable for scheme interpretation but remain a bottleneck for machine-verifiable specifications. 


\vspace{2mm}
\noindent\fbox{%
	\parbox{0.97\columnwidth}{%
		\textbf{Question. Can LLMs generate scheme models that are both interpretable and tool-executable?} Similar-looking outputs are only useful for verification when they satisfy formal tool constraints; in practice this still often needs guidance, repair, or hybrid pipelines.
    }
}

\subsection{\replaced[id=R2]{Case Studies of Evaluation Tasks}{The case study of evaluation tasks}} \label{sec:cases}

\begin{figure*}[t]
	\centering
	\includegraphics[width=0.8\textwidth]{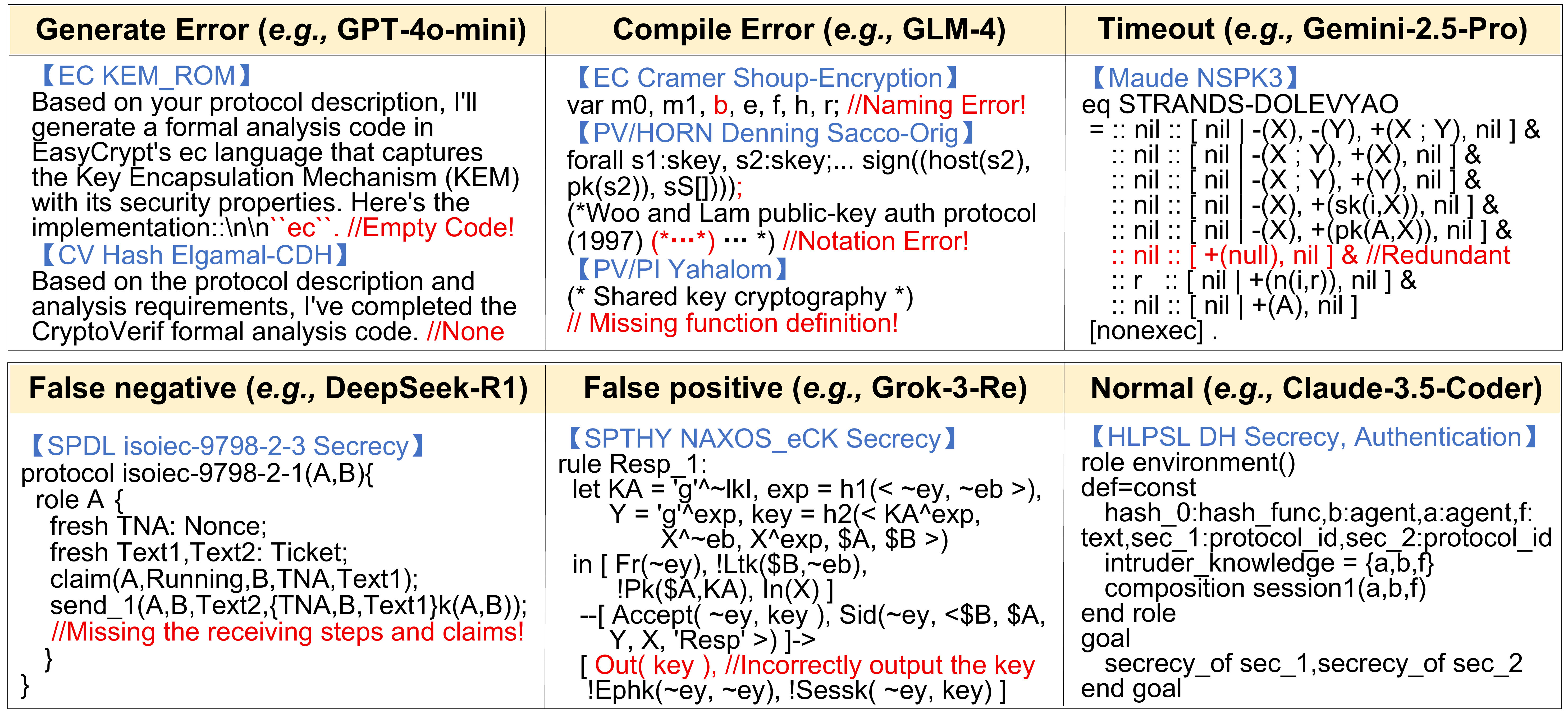}
	\caption{\added[id=R2]{Representative cases of LLM outputs across evaluation tasks, highlighting typical error patterns.}}
	\label{fig:cases}
	\vspace{-0.4cm}
\end{figure*}

Figure~\ref{fig:cases} summarizes representative LLM outputs across our five tasks and highlights five typical error patterns.\footnote{\added[id=R2]{Language-specific error modes are summarized on} \url{https://github.com/Secbrain/CrypFormBench/tree/main/experiments}.} \deleted[id=R2]{To better illustrate the challenges of formal code generation and analysis, we summarize representative cases of LLM outputs in Figure~\ref{fig:cases}, involving five typical error patterns.} 
\first \textit{\added[id=R2]{Generation errors.}} 
Models sometimes fail at the first step, producing either empty code or placeholder text (\eg EC of \textit{KEM\_ROM} scheme generated by \textit{GPT-4o-mini}), indicating weak grounding in the target syntax.
\second \textit{\added[id=R2]{Compilation errors.}} 
Even with near-correct syntax, outputs may still contain name conflicts, missing declarations, or misused keywords (\eg EC of \textit{Cramer-Shoup} encryption scheme completed by \textit{GLM-4}), where minor deviations break executability.
\third \textit{\added[id=R2]{Timeouts.}}  
Syntactically valid files can still diverge during analysis due to redundant or ill-structured rules (\eg MAUDE of \textit{NSPK3} protocol corrected by \textit{Gemini-2.5-Pro}), reflecting that model-generated code may introduce hidden inefficiencies that degrade analysis performance. 
\fourth \textit{\added[id=R2]{False negatives.}}  
Under-approximating the adversary or omitting critical claims can cause insecure schemes to be reported as \texttt{SAFE} (\eg SPDL of \textit{ISO/IEC-9798} transformed by \textit{DeepSeek-R1}), highlighting the importance of scheme process consistency. 
\fifth \textit{\added[id=R2]{False positives.}} 
Conversely, models may generate spurious outputs that violate claims (\eg SPTHY of \textit{NAXOS\_eCK} protocol completed by Grok-3), which may mislead users into believing a secure scheme is vulnerable. 


In contrast, strong code-oriented models (\eg \textit{Claude-3.5-Sonnet-Coder}) can produce valid specifications, correctly capturing both secrecy and authentication goals in HLPSL, demonstrating the potential of LLMs when guided properly. 
These findings reflect that existing LLMs are partially aware of the writing conventions of formal languages but fall short of producing fully correct, tool-compatible specifications. 
Nevertheless, such outputs are not \replaced{worthless}{without value}: with minor manual corrections, many of them can be repaired into valid specifications. 
This observation motivates three directions: \first employing \emph{refined} and \emph{few-shot} prompting to guide models toward stricter formal syntax, \second \emph{Pass@K} generation to increase the probability of obtaining a correct specification, and \third \textit{post-processing} to reduce trivial errors. 
These strategies are further discussed in \S~\ref{sec:optimizations}. 

\vspace{2mm}
\noindent\fbox{%
	\parbox{0.97\columnwidth}{%
		\textbf{Are LLMs reliable enough for direct verification?} Generated models often contain syntactic or semantic errors in one pass. Thus, it is suggested to treat their outputs as drafts, and apply strategies such as \emph{few-shot} guidance and \emph{Pass@K} to improve reliability before tool verification. 
	}
}

\subsection{\added{The Execution Efficiency}} \label{sec:efficiency}

\begin{figure*}[t]
	\centering
	\includegraphics[width=0.995\textwidth]{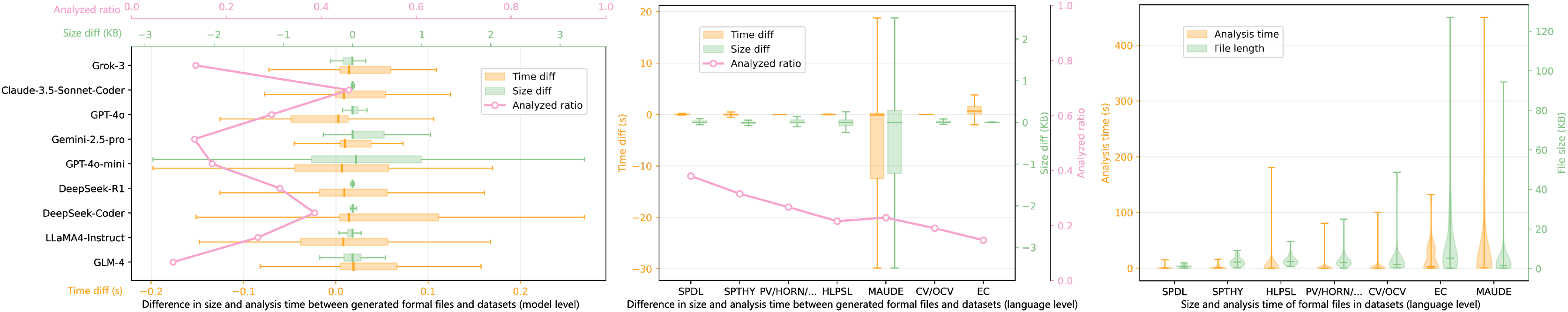}
	\caption{\added{Execution efficiency (analysis time and file size) of successfully analyzed files.}}
	\label{fig:timeuse}
	\vspace{-0.3cm}
\end{figure*}


Besides correctness, we evaluate \emph{execution efficiency} using the verifier runtime and file size on \emph{successfully analyzed} outputs. Figure~\ref{fig:timeuse} reports model-/language- level differences between generated and reference formal files, as well as baseline time/size distributions of the benchmark. 

\textit{Intrinsic cost of different verifiers (dataset baseline).}
\added{The right panel shows distinct runtime profiles across verifier languages. Lightweight tools such as Scyther and Tamarin usually finish within tens of seconds (mostly below $\sim$20s), whereas proof-/rewriting-heavy backends have heavy tails: AVISPA reaches $\sim$180s, and Maude-NPA/EasyCrypt exceed $\sim$400s. File length follows similarly, with EC up to $\sim$120KB and Maude up to $\sim$90KB, indicating higher complexity and overhead.}

\textit{Extra cost induced by LLM analyzable outputs.}
\added{For most models and languages, median analysis-time differences stay near 0 (roughly within $\pm$0.1-0.2s), and file-size differences are small (typically a few KB), suggesting that executable LLM outputs are usually comparable to human-curated scripts. Still, redundant declarations, over-nested terms, or ill-structured rules can inflate size and slow verification, causing timeouts even after compilation. Thus, efficiency is mainly dominated by intrinsic tool/language difficulty, while analyzability remains the key bottleneck for hard backends; it also indicates whether models learn verifier-friendly structure rather than merely valid syntax.}

\added{For most models and languages, median analysis-time differences stay near 0 (within $\pm$0.1-0.2s), and file-size differences are small (typically a few KB), suggesting executable LLM outputs are usually comparable to human-curated scripts. However, redundant declarations, over-nested terms, or ill-structured rules can inflate size and cause timeouts even after compilation. Thus, efficiency is mainly governed by intrinsic tool/language difficulty, while analyzability remains the key bottleneck. It also reveals whether models learn verifier-friendly structure beyond valid syntax.}


\vspace{2mm}
\noindent\fbox{%
  \parbox{0.97\columnwidth}{%
    \textbf{\added{Question. Do analyzable LLM-generated scripts run efficiently on formal verifiers?}}
  Efficiency differs sharply by backend: lightweight symbolic tools (\eg Scyther and Tamarin) usually complete within $\sim$20s, while complex tools (\eg EasyCrypt and Maude-NPA) may exceed $\sim$400s and remain timeout-prone. In practice, reliable workflows should combine guided prompting/repair with timeout-aware constraints to keep outputs both executable and efficient. 
  }
}

	\section{Discussion and Suggestions} \label{sec:discussion}

We summarize several insights and practical recommendations derived from our experiments, focusing on prompt strategies, model training, evaluation mechanisms, and security extensions.

\subsection{Sensitivity for Tunable Parameters in Metric Design} \label{sec:parameters}
\begin{figure*}[t]
\begin{minipage}{0.54\textwidth}
	\centering
	\setlength\tabcolsep{3 pt}
	\renewcommand{\arraystretch}{1.25}
	\captionof{table}{Alternative scoring configurations (C1-C4).}
	\label{table:sens_configs}
	\resizebox{1.0\linewidth}{!}{
	\begin{tabular}{cccccc cccc c}
		\toprule
		Config &
		$w_{\text{gen}}$ & $w_{\text{comp}}$ & $w_{\text{trans}}$ & $w_{\text{corr}}$ & $w_{\text{interp}}$ &
		$\alpha$ & $\beta$ & $\lambda_{\text{err}}$ & $\lambda_{\text{false}}$ &
		$\gamma$ \\
		\midrule
		C1 & 0.25 & 0.20 & 0.25 & 0.15 & 0.15 & 0.30 & 0.30 & 0.40 & 0.60 & 1.0 \\
		C2 & 0.20 & 0.20 & 0.20 & 0.20 & 0.20 & 0.30 & 0.30 & 0.40 & 0.60 & 1.0 \\
		C3 & 0.30 & 0.15 & 0.30 & 0.10 & 0.15 & 0.30 & 0.30 & 0.40 & 0.60 & 1.0 \\
		C4 & 0.25 & 0.20 & 0.20 & 0.15 & 0.20 & 0.40 & 0.20 & 0.50 & 0.50 & 0.5 \\
		\bottomrule
	\end{tabular}}
\end{minipage}
\hfill
\begin{minipage}{0.43\textwidth}
	\centering
	\includegraphics[width=\textwidth]{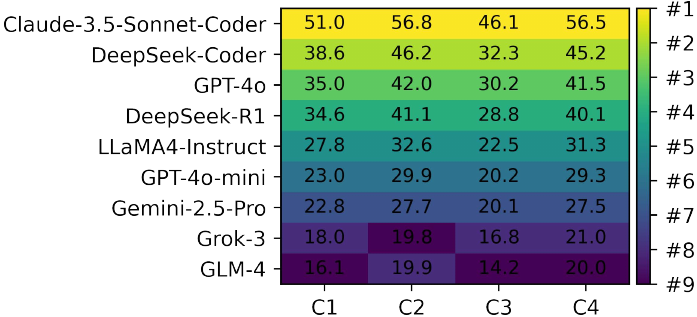}
	\captionof{figure}{Overall ranks of LLMs under C1-C4.}
	\label{fig:sens_heatmap}
\end{minipage}
\vspace{-0.3cm}
\end{figure*}

To validate that our conclusions are not artifacts of a particular setting, we evaluate four alternative parameter configurations shown in Table~\ref{table:sens_configs}, covering (C1) default difficulty task weights, (C2) uniform weights, (C3) heavier weights on generation/transformation, and (C4) modified sub-weights of interpretation and correction with a weaker analyzability penalty. 
As shown in Fig.~\ref{fig:sens_heatmap}, across these settings, model rankings are highly stable: the top models remain unchanged, and only a swap between the two lowest-ranked models (GLM-4/Grok-3) under uniform weights. 
Also, C1 has a high ranking correlation with other settings (Spearman $\rho \ge 0.98$ and Kendall $\tau \ge 0.94$), indicating that our main findings are robust to reasonable variations of tunable parameters.

\subsection{Validation of Embedding-Based Similarity}
For interpretation, we use embedding-based cosine similarity ($s_{\text{logic}}$ and $s_{\text{anno}}$) for logic descriptions and code annotations (\cf \S~\ref{sec:metrics}), and validate it from two aspects.
\first \textit{\added[id=R2]{Cross-encoder robustness.}}
Besides the default \textit{Qwen3-Embedding-8B}, we recompute all scores with \textit{BGE-large}~\cite{bge} and \textit{E5-large-v2}~\cite{e5}. Over 1,260 interpretation instances, Qwen3 shows strong agreement with BGE/E5, with Pearson correlation $\ge 0.98$ and Spearman correlation $\ge 0.88$, indicating that model comparisons and rankings are insensitive to encoder choice. Table~\ref{tab:interp_audit} further demonstrates case-level consistency.
\second \textit{Human audit on high-/low-similarity cases.}
We identify 711 high-similarity ($\ge 0.9$) and 40 low-similarity ($\le 0.3$) instances, and manually inspect 20 random cases from each group. High-similarity outputs generally preserve roles, message flows, and security goals, supporting valid formal-model reconstruction, whereas low-similarity outputs often miss or mis-specify key semantics, such as secrecy goals, nonce bindings, or adversary assumptions. Table~\ref{tab:interp_audit} gives examples.\footnote{Additional rare cases are available at \url{https://github.com/Secbrain/CrypFormBench/tree/main/examples}.} Rare divergences, \eg nsl3.spdl, show that annotation similarity alone can be misleading, motivating our multi-signal interpretation score that combines logic similarity, annotation similarity, and tool-executability/verification outcomes (\cf Eq.~(\ref{eq:interpretation})). The natural-language references derived from formal specifications were also manually reviewed during benchmark construction, while MT-Bench/Chatbot Arena~\cite{NeurIPS_MT-Bench} suggests LLM-as-judge as an additional future signal.


\begin{table*}[t]
	\centering
	\setlength\tabcolsep{1.5 pt}
	\scriptsize
	\caption{Representative high-/low-consensus interpretation cases for human audit.}
	\label{tab:interp_audit}
	\resizebox{1.0\linewidth}{!}{
	\begin{tabular}{lllcccc|lllcccc}
		\toprule
		\textbf{\added{Aspect}} & \textbf{\added{File}} & \textbf{\added{LLM}} & \textbf{\added{Qwen3}} & \textbf{\added{BGE}} & \textbf{\added{E5}} & \textbf{\added{Human}} & \textbf{\added{Aspect}} & \textbf{\added{File}} & \textbf{\added{LLM}} & \textbf{\added{Qwen3}} & \textbf{\added{BGE}} & \textbf{\added{E5}} & \textbf{\added{Human}} \\
		\midrule
		\added{Notation} & andrew-lowe-ban.spdl & \added{Claude-3.5} & \added{0.98} & \added{0.96} & \added{0.982} & \added{Correct} & \added{Logic}    & woo-lam.spdl         & \added{LLaMA4}     & \added{0.02} & \added{0.022}& \added{0.019} & \added{Incorrect} \\
		\added{Logic}    & TLS-PSK.spthy        & \added{GPT-4o}     & \added{0.96} & \added{0.94} & \added{0.95}  & \added{Correct} & \added{Notation} & OtwayRees.pv         & \added{LLaMA4}     & \added{0.12} & \added{0.18} & \added{0.15}  & \added{Miss secrecy} \\
		\added{Logic}    & IKEv2-DS.hlpsl       & \added{GPT-4o}     & \added{0.96} & \added{0.97} & \added{0.96}  & \added{Correct} & \added{Logic}    & Anonymous.hlpsl & \added{Grok-3}  & \added{0.38} & \added{0.361}& \added{0.364} & \added{Incorrect} \\
		\added{Notation} & IKEv2-DS.hlpsl       & \added{GPT-4o}     & \added{0.98} & \added{0.97} & \added{0.97}  & \added{Correct} & \added{Notation} & Anonymous.hlpsl & \added{Grok-3}  & \added{0.01} & \added{0.02} & \added{0.01}  & \added{Miss code} \\
		\added{Notation} & nsl3.spdl            & \added{LLaMA4}     & \added{0.975}& \added{0.981}& \added{0.983} & \added{Correct} & \added{Logic}    & nsl3.spdl            & \added{LLaMA4}     & \added{0.019}& \added{0.027}& \added{0.022} & \added{Incorrect} \\		
		\bottomrule
	\end{tabular}}
	\vspace{-0.3cm}
\end{table*}

\subsection{\added{Few-Shot and Fine-Tuning Optimizations}} \label{sec:optimizations}

\begin{figure*}[t]
	\centering
	\includegraphics[width=0.995\textwidth]{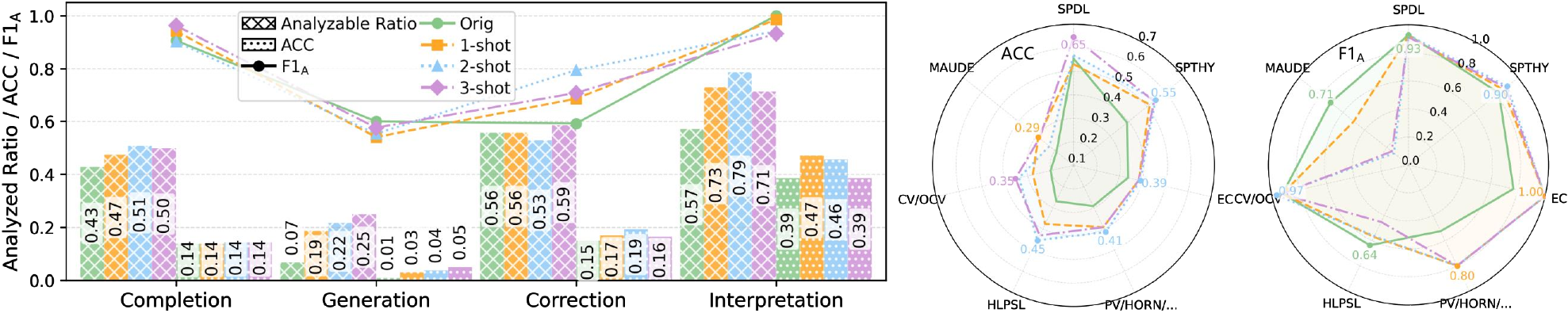}
	\caption{\added{Effects of $K$-shot prompting strategy ($K\in\{0,1,2,3\}$) in task and language perspectives. Transformation is omitted since it does not yield analyzable outputs for all $K$ in this experiment.}}
	\label{fig:fewshot}
	\vspace{-0.3cm}
\end{figure*}


\textit{\added{Few-shot strategy.}} 
We evaluate $K$-shot prompting ($K\in\{0,1,2,3\}$) as a simple retrieval-augmented strategy~\cite{NeurIPS_RAG,DASFAA_CyberLLM} on a 10\% dataset subset, averaging results across models. As shown in Fig.~\ref{fig:fewshot}, few-shot markedly improves analyzability for generation (6.6\% at 0-shot $\rightarrow$ 24.9\% at 3-shot) and moderately for completion (42.9\% $\rightarrow$ 50.9\%) while keeping ACC $>92\%$. For correction and interpretation, analyzability is stable, but ACC increases notably (\eg 55.6\% $\rightarrow$ 72.9\% at 3-shot). Exemplars mainly help reproduce tool-specific scaffolding, \eg imports/modules, role/environment blocks, and event/query declarations, improving parsing and execution. Remaining failures are mostly semantic or tool-specific, such as Maude-NPA timeouts from divergent rewriting and EasyCrypt/CryptoVerif failures from inconsistent game transitions or event definitions. Thus, few-shot improves analyzability and correctness, but does not close the gap between easy tasks (interpretation/completion) and hard tasks (generation/transformation)~\cite{brown2020language,wei2022emergent}.

\begin{figure*}[t]
	\centering
	\includegraphics[width=0.995\textwidth]{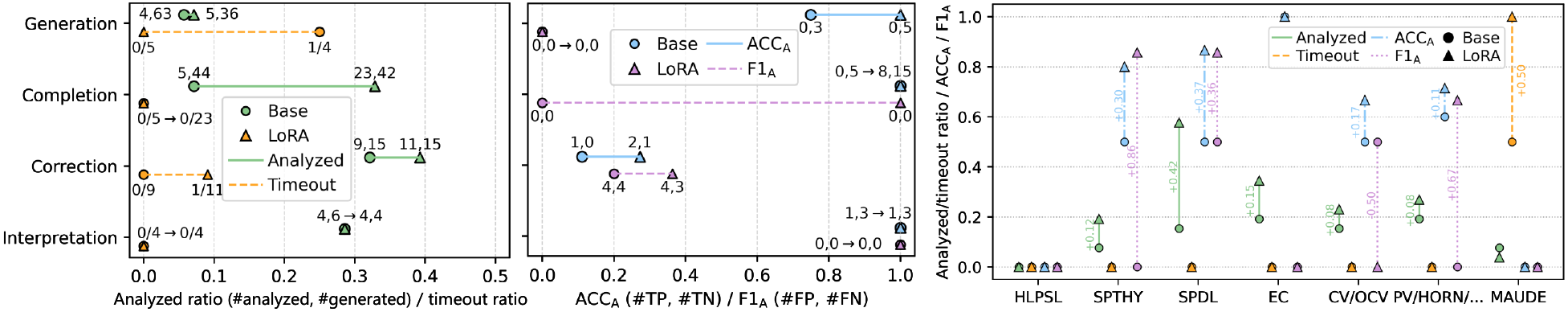}
	\caption{Effects of LoRA fine-tuning on \textit{Qwen2.5-Coder-3B} from task- and language-level views. The timeout ratio is computed as \#timeout/\#analyzed, and the transformation is omitted as it has no analyzable outputs.}
	\label{fig:lora}
	\vspace{-0.2cm}
\end{figure*}

\paragraph{\added{LLM Fine-tuning.}} 
We fine-tune \textit{Qwen2.5-Coder-3B}~\cite{Qwen25-Coder} with LoRA to assess task-specific model improvement. For each task and verifier language, we use a 90\%/10\% train/evaluation split and train for 3 epochs. Figure~\ref{fig:lora} reports task- and language-level changes in analyzability, timeout ratio, and correctness on analyzable outputs (ACC$_\text{A}$/F1$_\text{A}$). LoRA substantially improves completion (analyzable outputs 5$\rightarrow$23, TP 0$\rightarrow$8, TN 0$\rightarrow$15) and modestly improves correction (analyzable 9$\rightarrow$11, TP 1$\rightarrow$2, TN 0$\rightarrow$1, FN 4$\rightarrow$3). Generation only slightly improves in analyzability (4$\rightarrow$5), interpretation is largely unchanged, and transformation still yields zero analyzable outputs. We also observe language-dependent gains: LoRA helps easier languages such as SPDL, but remains limited for harder ones such as HLPSL and EC, where failures stem from tool-specific constraints, \eg divergence/timeouts or proof inconsistencies. task-specific LoRA can improve analyzability and correctness for completion (and to a lesser extent correction), while the difficulty ordering remains stable: interpretation/completion is much easier than generation/transformation.

\begin{figure*}[t]
	\centering
	\includegraphics[width=0.995\textwidth]{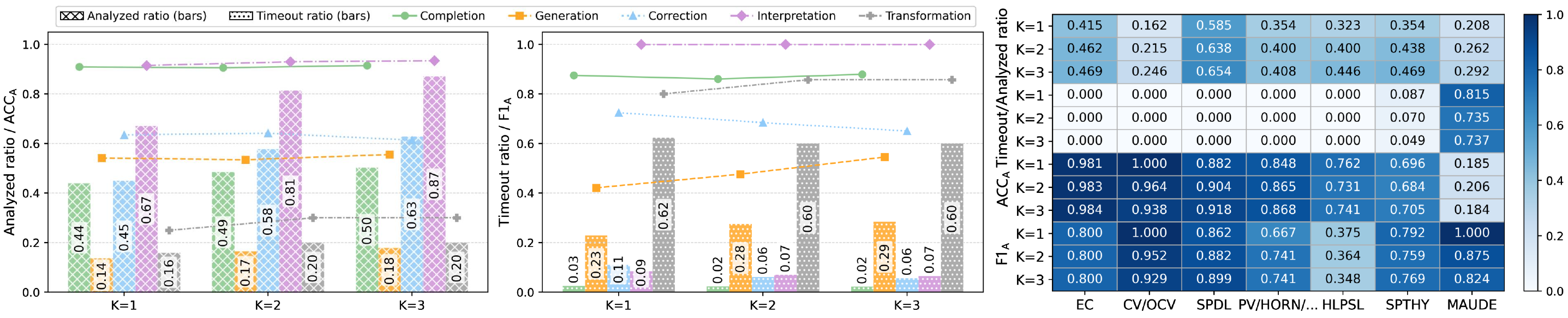}
	\caption{Effects of PASS@K ($K\!=\!1,2,3$) mechanism on task- and language-levels.}
	\label{fig:passk}
	\vspace{-0.3cm}
\end{figure*}

\paragraph{\added{PASS@K mechanism.}} 
We query the LLM up to $K$ times per instance, verify each candidate, and aggregate the best outcome: an instance is analyzable if any attempt is analyzable, and correctness is computed on this analyzable set~\cite{chen2021evaluating}. Figure~\ref{fig:passk} shows task- and language-level changes as $K$ increases from 1 to 3. Larger $K$ consistently improves the analyzed ratio, especially for hard tasks such as generation and transformation, where single-shot outputs often fail parsing or verification. Timeout ratios usually decrease, though some later attempts reach the verifier but trigger expensive or non-terminating analyses. On analyzable outputs, ACC$_\mathrm{A}$ and F1$_\mathrm{A}$ remain mostly stable or improve mildly, indicating that Pass@K mainly alleviates the executability bottleneck rather than improving already-analyzable candidate quality. Language-level results show larger analyzability gains for stricter or more brittle pipelines, and higher timeout growth for timeout-prone backends. Thus, Pass@K is a practical complement to single-shot evaluation, but some candidates remain semantically ill-posed even after becoming analyzable.

\paragraph{Other optimizations.} 
Beyond the above methods, promising directions include \textit{structured prompts} that separate roles/assumptions/goals to reduce ambiguity, \textit{multilingual prompting} to better align natural reasoning with formal syntax~\cite{xu2023multilingual}, and \textit{Chain-of-Thought (CoT) prompting}~\cite{wei2022chain} to encourage step-wise decomposition of verification logic. Also, \textit{adaptive tool invocation} for backends, where LLMs synthesize tool-specific run commands from scheme logic and target properties, especially for non-default adversary settings such as KCI or (weak) perfect forward secrecy (wPFS/PFS). 



\subsection{Fine-Grained Property-Level Evaluation} \label{sec:properties}
\begin{figure*}[t]
	\centering
	\includegraphics[width=0.995\textwidth]{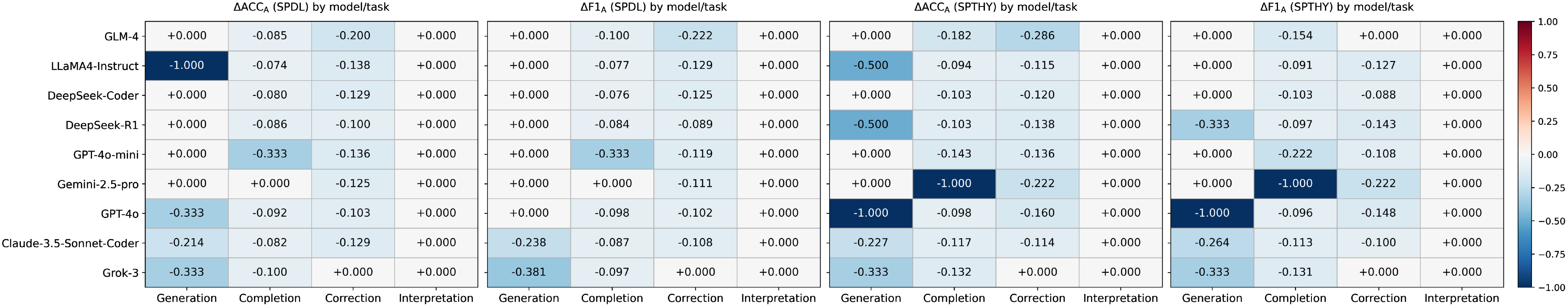}
	\caption{\added[id=R1]{Performance of LLMs on fine-grained (property-level) evaluation, where $\Delta = \text{Fine} - \text{Coarse}$.}}
	\label{fig:finelevel}
	\vspace{-0.1cm}
\end{figure*}

Although property-level evaluation is important~\cite{cortier2018survey}, \fram uses a \emph{scheme-level} label: a scheme is \texttt{UNSAFE} if any checked goal is violated, and \texttt{SAFE} otherwise. This supports uniform evaluation across heterogeneous tools and end-to-end usability. Since each instance typically has 2-4 goals (\S~\ref{sec:dataset}), fully automatic property-level scoring is difficult for tools with only summary verdicts (\eg AVISPA) or requiring trace/proof parsing to attribute failures to goals (Maude-NPA, EasyCrypt, and CryptoVerif). We therefore analyze SPDL and SPTHY, whose verifiers expose clear per-goal outcomes. Each (scheme, goal) pair is a judgment: TP if the goal is correctly verified, FP if a satisfied goal is spuriously reported as \texttt{UNSAFE}, FN if an expected goal is missing or incorrect, and TN if a \texttt{SAFE} goal is correctly verified. A scheme is fully correct only when all goals are covered and correct. 
Figure~\ref{fig:finelevel} reports $\Delta=\text{Fine}-\text{Coarse}$ for ACC$_\mathrm{A}$ and F1$_\mathrm{A}$. Generation drops most due to low analyzability and missing/mis-modeled goals. For completion and correction, the reduction is moderate (about 10\% on average), while interpretation is nearly unchanged. Model rankings remain highly stable, showing that our conclusions are insensitive to aggregation granularity. Extending property-level evaluation to all seven verifiers requires dataset refactoring, \eg splitting multi-goal scripts, and tool-specific parsers for per-goal verdicts, which we leave as future work.

\subsection{\added{Other Considerations}}


\textit{\added[id=R2]{Improving analyzability bottlenecks.}} 
Low analyzability mainly comes from tool-level failures (parse/type errors) and timeouts, especially in transformation. Non-analyzable outputs often miss mandatory scaffolding (imports/roles/queries) or contain typing/scope mismatches, while analyzable-but-wrong cases usually mis-encode adversaries/goals, \eg freshness, binding, and correspondence. Promising directions include: \first \textit{grammar-constrained decoding} or \textit{schema-guided generation} to enforce target-language structure and avoid mixed-language fragments. \second \textit{Two-stage translation} via a tool-agnostic IR of roles, message flows, and goals. \third \textit{Tool-in-the-loop repair} using parser/type-checker errors and verifier diagnostics, optionally with few-shot scaffolding examples.

\paragraph{\added[id=R2]{Inconsistent results of various verifiers.}} 
Verifiers may disagree on the same scheme due to different abstractions, supported primitives, and threat/property models. For example, AVISPA can detect the UKS attack of STS, while tools such as Scyther cannot faithfully encode the required adversary behaviors. In \fram, we resolve such conflicts using canonical labels from cryptographic literature/standards under specified threat models (\S~\ref{sec:collection}). Although resolving cross-tool disagreement is beyond our scope, \fram enables future study through transformation tasks and the multi-tool harness, which collect verdicts and attack traces to analyze whether disagreements arise from modeling choices, property definitions, or semantic gaps. False reports may also be reduced by LLM-assisted attack-path verification with explicit attack scenarios.

\begin{figure*}[t]
	\centering
	\includegraphics[width=0.995\textwidth]{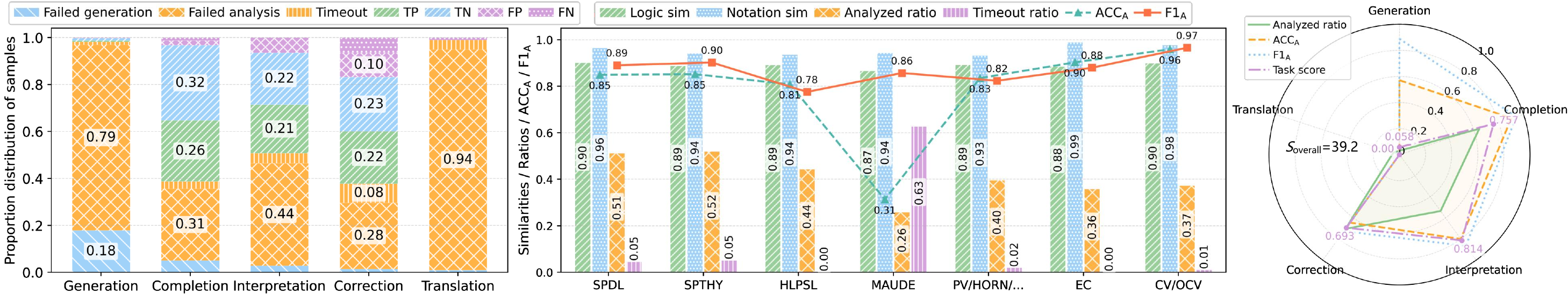}
	\caption{Performance of GPT-5.1 on language and task levels.}
	\label{fig:gpt5}
	\vspace{-0.3cm}
\end{figure*}

\paragraph{\added{Scalability of model and dimension.}} 
We evaluate the newly released \textit{GPT-5.1}~\cite{gpt51} to show that \fram can incorporate emerging LLMs without changing the pipeline. Figure~\ref{fig:gpt5} reports its task- and language-level results, including execution outcomes, interpretation similarity, and $S_{\text{overall}}$. Compared with \textit{GPT-4o-mini} and \textit{GPT-4o}, \textit{GPT-5.1} improves the overall score (23.0$\rightarrow$35.0$\rightarrow$39.2), likely due to stronger instruction following and adaptive reasoning, but keeps a similar profile: interpretation/completion are strong, while generation/transformation remain limited by analyzability and timeouts. Beyond model scaling, the same framework can be extended to more verifiers (\eg Verifpal~\cite{INDOCRYPT_Verifpal} and SAPIC+~\cite{USENIX_SAPIC+}) and richer protocol families (\eg 6G and blockchain~\cite{ICWS_Contract,ICAIS_Key}), improving coverage and realism.


\paragraph{\added[id=R3]{Model contamination.}}
\fram includes standard schemes (\eg Needham-Schroeder, TLS/SSH-style handshakes, and 5G-AKA) whose descriptions or partial formalizations may appear online, so we cannot guarantee zero pre-training overlap. However, exposure to high-level schemes is unlikely to solve our tasks: transformation and correction require tool- and language-specific reasoning. As experimented in \S~\ref{sec:evaluation}, models perform well on interpretation/completion but still struggle with generation/transformation across tools. This consistent pattern across open/closed models suggests that \fram mainly probes genuine capability gaps rather than memorized code.

	\section{Related Work}
\subsection{Program Synthesis and Symbolic Models}
Symbolic models can be viewed as a domain-specific form of programs. Prior work on program synthesis has studied translating natural language into executable artifacts such as SQL queries~\cite{ACL_SQL}, bash commands~\cite{Washington_synthesis}, or general-purpose code~\cite{Science_alphacode, chen2021evaluating, JMLR_PaLM, ICLR_CodeGen}. These advances demonstrate that LLMs can generate structured outputs across domains~\cite{CAV_Enumerative}. 
However, symbolic specifications differ fundamentally from SQL or Python. They demand strict grammar, security-critical semantics, and high precision, while training data is extremely scarce. Multi-stage pipelines for scheme modeling have been proposed~\cite{ICSE_Protocol}, yet systematic evaluation of LLMs on formal security protocols is still missing. Our work addresses this gap by constructing a large-scale benchmark, enabling standardized and reproducible evaluation of LLMs on symbolic protocol analysis tasks.

\subsection{LLM-Aided Formal Verification}
Formal verification has long been used to analyze security properties such as authentication, secrecy, and replay resistance. Symbolic models (\eg Dolev-Yao) capture message flows, while computational proofs provide stronger guarantees under cryptographic assumptions. Recent work applies LLMs to verification-related tasks, including temporal-logic generation from natural language~\cite{EMNLP_NL2TL}, fine-tuned specification synthesis~\cite{JMLR_Transformer}, interactive correction~\cite{CAV_nl2spec,OOPSLA_Interactive,ICSE_Protocol}, and proof guidance with Coq/Lean~\cite{CoRR_Theorem,ACL_DTSolver,NeurIPS_LeanDojo}. However, these studies mainly target logic synthesis or proof assistance. We instead evaluate end-to-end LLM capabilities across five tasks and seven formal tools, bridging natural-language scheme descriptions and automated verification.

	\section{Conclusion and Future Work}

This paper presents the first comprehensive and well-designed benchmark \fram for formal cryptographic scheme analysis supporting both symbolic and computational security. \fram establishes a unified evaluation framework and comprises 700 cryptographic instances covering 677 distinct schemes and primitives, encompassing both classical and modern designs, as well as simple and complex cases. It supports 160 security property verifications and enables systematic measurement of five LLM capabilities in formal cryptographic scheme analysis. Also, \fram incorporates a multi-dimensional scoring mechanism, offering a comprehensive evaluation of mainstream LLMs. In the future, we will extend \fram with more verifiers and schemes, and develop a verifier-in-the-loop and agentic LLM framework that integrates the explored strategies for tool-usable formal analysis. 


	
    
    \section{Data Availability}
    
    We release both the source code and the curated dataset of \fram at our online repository \cite{fram}.

    \section*{Acknowledgements}
    This work was partially funded by the National Cryptologic Science Fund of China (2025NCSF01008) and the fund of Laboratory for Advanced Computing and Intelligence Engineering.

    \clearpage
    
	\bibliographystyle{ACM-Reference-Format}
	\bibliography{ref}

\end{document}